\documentclass[preprint, times,12pt]{elsarticle}

\usepackage{amssymb}
\usepackage{amsmath}
\usepackage{booktabs}
\usepackage{multirow}
\usepackage{longtable,tabularx}
\usepackage{threeparttable}
\usepackage{url}

\usepackage{subcaption}
\usepackage[margin=1in]{geometry}

\journal{Transportation Research Part A: Policy and Practice}

\begin{document}

\begin{frontmatter}



\title{Examining the Dynamics of Local and Transfer Passenger Share Patterns in Air Transportation}


\author[1]{Xufang Zheng} %% Author name
\ead{xufang.zheng@aa.com}

\author[2]{Qilei Zhang\corref{cor1}
%\fnref{fn1}
} %% Author name
\ead{q.zhang@cqu.edu.au}

\author[3]{Victoria Cobb}
\ead{vcobb@umich.edu}

\author[3]{Max Z. Li} %% Author name
\ead{maxzli@umich.edu}

\cortext[cor1]{Corresponding author}

\affiliation[1]{organization={American Airlines},%Department and Organization
            addressline={1 Skyview Dr},  
            city={Fort Worth},
            postcode={76155}, 
            state={Texas},
            country={United States}}

\affiliation[2]{organization={Central Queensland University},%Department and Organization
            addressline={42-52 Abbott Street \& Shields Street}, 
            city={Cairns},
            postcode={4870}, 
            state={Queensland},
            country={Australia}}

\affiliation[3]{organization={University of Michigan},%Department and Organization
            addressline={1320 Beal Ave}, 
            city={Ann Arbor},
            postcode={48109}, 
            state={Michigan},
            country={United States}}

%% Abstract
\begin{abstract}
The air transportation local share, defined as the proportion of local passengers relative to total passengers, serves as a critical metric reflecting how economic growth, carrier strategies, and market forces jointly influence demand composition. This metric is particularly useful for examining industry structure changes and large-scale disruptive events such as the COVID-19 pandemic. This research offers an in-depth analysis of local share patterns on more than 3900 Origin and Destination (O\&D) pairs across the U.S. air transportation system, revealing how economic expansion, the emergence of low-cost carriers (LCCs), and strategic shifts by legacy carriers have collectively elevated local share. To efficiently identify the local share characteristics of thousands of O\&Ds and to categorize the O\&Ds that have the same behavior, a range of time series clustering methods were used. Evaluation using visualization, performance metrics, and case-based examination highlighted distinct patterns and trends, from magnitude-based stratification to trend-based groupings. The analysis also identified pattern commonalities within O\&D pairs, suggesting that macro-level forces (e.g., economic cycles, changing demographics, or disruptions such as COVID-19) can synchronize changes between disparate markets. These insights set the stage for predictive modeling of local share, guiding airline network planning and infrastructure investments. This study combines quantitative analysis with flexible clustering to help stakeholders anticipate market shifts, optimize resource allocation strategies, and strengthen the air transportation system's resilience and competitiveness.
\end{abstract}


% \begin{highlights}
% \end{highlights}

%% Keywords

\end{frontmatter}



%%%%%%%%%%%%%% Our Text %%%%%%%%%%%%%%%%%

\section{Introduction}\label{sec:Intro}

    \subsection{Background}\label{sec:Intro_BG}
    
    Air transportation is a pivotal component of the global economy, facilitating the movement of passengers and goods. The U.S. commercial air carrier industry experienced a decade of relative stability from the end of the Great Recession in 2009 until the outbreak of COVID-19 in 2020~\cite{FAAAerospaceForecast}. The COVID-19 pandemic brought unparalleled challenges, leading to drastic reductions in passenger numbers, financial losses, and operational disruptions~\cite{ACIImpactofCOVID19}. In 2020, several small U.S. regional carriers ceased operations. Later, each passenger carrier across the industry continued to face generally similar headwinds and tailwinds as the others~\cite{bloom2020airlines}. Demand for travel to leisure domestic destinations and specific regions (e.g., Latin America) surged. Carriers were caught off guard and struggled to bring aircraft back into service, open new routes, and hire enough staff to meet the increased levels of demand. In 2023, the landscape transformed again. As a wider array of accessible destinations opened up, travelers responded by seeking flights across the Atlantic and to some Pacific markets~\cite{FAAAerospaceForecast}. The recovery in patterns is now driving more structural changes, reflecting a new reality for the aviation market~\cite{ACIAirportIndustry}. As the industry rapidly and steadily recovered, there was a renewed focus shift towards adapting to new realities and reshaping strategies for enhancing industry resilience and sustainability. One particular area of interest is the research of passenger flows, specifically the dynamics of local and transfer passengers during the recovery after the pandemic's impact.  
    
    Passengers traveling through an airport can be divided into local passengers and transfer (or connecting) passengers. Taking A $\rightarrow$ B as a directional O\&D pair example, local passengers on A $\rightarrow$ B are the passengers originally flying out from A to B via direct flight services. Transfer passengers on A $\rightarrow$ B are the passengers originally flying out from other airports but taking A as the last transfer stop to B. Both local and transfer passengers take flights from airport A to B, however, the passenger behaviors (e.g., local transportation choice), services needed (e.g., security check and customer service) and impact on the airport (e.g., passengers moving flow at the airport) are different for the two groups of passengers. Therefore, the volumes and shares of local and transfer passengers are useful metrics and decision-making considerations for different stakeholders in air transportation~\cite{DevelopmentoTransferPassenger}. Uncovering trends and characteristics regarding the share of local and transfer passengers is important for providing valuable insights into airport operations, passenger behavior, and strategic planning~\cite{ACIAirportIndustry}.

    \subsection{Problem Significance}\label{sec:Intro_PS}
    
        % - Local Share analysis - important in for airlines 
        \subsubsection{Significance to Airlines}
        Passenger demand and itinerary preference are key concerns in airline management and operations, including marketing strategy, network planning, customer service, etc. Different types of airlines employ distinct business models. Legacy carriers, or Full-Service Carriers (FSCs), traditionally operate hub-and-spoke (H\&S) networks, linking a set of spoke airports to a central hub to efficiently route transfer passengers~\cite{AnIntegratedConnectionPlanning}. This network configuration consolidates the demand and reduces the need for numerous direct routes, thus achieving economies of scale. In contrast, Low-Cost Carriers (LCCs) predominantly utilize a more point-to-point (P2P) configuration, reducing the operational overhead of complex hub operations and focusing on direct services~\cite{AnIntegratedConnectionPlanning}. Ultra Low-Cost Carriers (ULCCs) further emphasize P2P operations, offering lower base fares with minimal included services, thus appealing to price-sensitive local passengers. FSCs can leverage their network complexity to accommodate greater volumes of transfer passengers, while LCCs and ULCCs primarily target O\&D markets with strong local passenger demand. Airlines can utilize insights from local and transfer passenger research to optimize their route structures, identify competitive niches, and align pricing and marketing strategies with underlying passenger segmentation. This enables airlines to attract more originating passengers through loyalty programs, partnerships, and promotional campaigns. 
        
        % - Local Share analysis - important in for airport 
        \subsubsection{Significance to Airports}
        Airports can benefit significantly from understanding local and transfer passenger dynamics. As local and transfer passengers exhibit distinct operational footprints, airport planning routinely assesses various infrastructure configurations and operational scenarios to accommodate differing proportions of local and transfer traffic~\cite{AirportSystemsPlanning}. Insights on local passenger shares allow airports to identify the proportion of originating travelers and distinguish them from all transiting passengers, including those on connecting itineraries. A high local passenger share indicates an airport's prominence as a primary departure point, necessitating efficient resource allocation and infrastructure development tailored to local market needs. For example, flights landing or departing from international hubs tend to carry a substantial volume of connecting passengers, leading to distinct resource requirements compared to airports dominated by local passengers~\cite{ForecastingAirportTransferPassenger}. More check-in counters, parking facilities, and departure lounges are often required to accommodate local passengers~\cite{ACIAnnualWorldAirport, DevelopmentoTransferPassenger}. On the other hand, transfer passengers rely heavily on more efficient terminal layouts, gate accessibility, and efficient intra-airport mobility options to ensure seamless connections~\cite{EvaluationofLevelofService, EvolutionofTheEuropean}. Understanding local share is pivotal in aligning infrastructure investment and operational management with passenger composition, thereby improving service quality and revenue streams. Originating passengers in particular contribute significantly to non-aeronautical income through parking fees, retail purchases, and ancillary services, while transfer passengers require streamlined terminal operations to ensure timely connections. These insights support nuanced market analysis, informed route development, and guided competitive benchmarking, thus enabling airports to enhance their competitive position. Airports can also leverage local share analysis to shape terminal design and gate allocations~\cite{OptimizationofTerminalAllocation}. Moreover, transfer passengers constitute a critical element in the intricate balance of airport costs and revenues in airport planning and management~\cite{AirlineHubbing—someImplications}, making it essential to benchmark local share against comparative peer airports to identify opportunities for growth.
        
        % - Local Share analysis - important in for FAA
        \subsubsection{Significance to Aviation Authorities}
        Insights into local and transfer passenger dynamics are also valuable for civil aviation authorities and air navigation service providers (ANSPs) such as the Federal Aviation Administration (FAA). These agencies make strategic decisions regarding airspace management, investment allocation, and infrastructure planning, all of which are enhanced by detailed passenger composition data. System-level fluctuations in the local and transfer shares can guide ANSPs in optimizing air traffic management protocols, capacity assessments, and long-term policy frameworks, promoting efficient system-wide operations. At the airport level, understanding the local versus transfer shares enables civil aviation authorities to anticipate infrastructure needs, streamline security procedures, and allocate resources to address peak demands. 

    % - Local Share analysis - what we want to achieve in this research
    \subsection{Contributions of Work}
    This research aims to clarify the dynamic interplay between local and transfer passengers and to establish a data-driven framework to identify local share patterns according to various market attributes and temporal evolutions. Particular emphasis is placed on examining pre- and post-pandemic changes, revealing large-scale disruptive events' impacts on air transportation networks, airport operations, and airline network configurations. By employing advanced time series clustering techniques, this research categorized O\&D pairs according to their local share characteristics, thereby uncovering patterns that have remained hidden in more aggregated approaches. The inclusion of robust clustering metrics further ensured that the derived groupings were both valid and meaningful, contributing to a more granular and actionable understanding of passenger flows within the air transportation network.
    
    % - definition of Local Share
    \subsection{Definitions}\label{sec:Intro_Def}
\begin{figure}[htbp]
        \centering
        \includegraphics[width=0.5\textwidth]{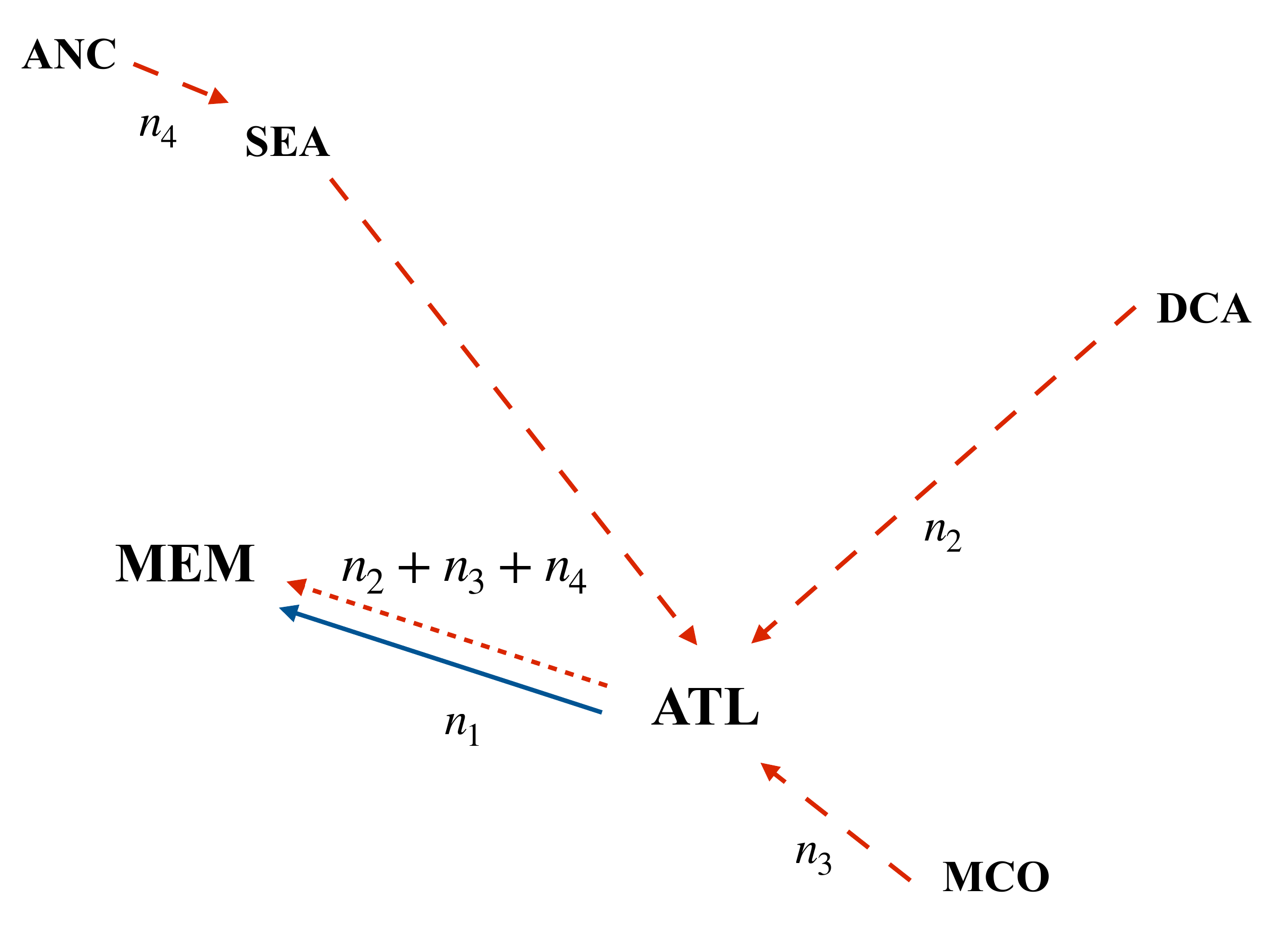}
        \caption{Simplified itineraries with last leg ATL $\rightarrow$ MEM.}
       \label{fig:LocalShare_def}
    \end{figure}

    Until now, we have provided a qualitative definition of local and transfer passenger shares. Here, we present explicit definitions for both. Take, for example, simplified itineraries that share a common final leg on the directional O\&D pair ATL (Hartsfield-Jackson Atlanta International Airport, Atlanta, GA) $\rightarrow$ MEM (Memphis International Airport, Memphis, TN) (Figure~\ref{fig:LocalShare_def}). There are four itineraries taking passengers to MEM through ATL, which are ATL $\rightarrow$ MEM, ANC (Ted Stevens Anchorage International Airport, Anchorage, AK) $\rightarrow$ SEA (Seattle-Tacoma International Airport, Seattle, WA) $\rightarrow$ ATL $\rightarrow$ MEM, DCA (Ronald Reagan Washington National Airport, Arlington, VA) $\rightarrow$ ATL $\rightarrow$ MEM, and MCO (Orlando International Airport, Orlando, FL) $\rightarrow$ ATL $\rightarrow$ MEM. Passengers originally departing from ATL to MEM on direct flights during time $t$ are local passengers on ATL $\rightarrow$ MEM, denoted as $n_1$. In contrast, passengers using ATL as the last transfer stop during the same time are transfer passengers, which include $n_2$, $n_3$, and $n_4$. The metric of local share quantifies the proportion of passengers originating from an airport relative to its total passenger volume, as defined in Eq.~\eqref{eq:localShare}. Correspondingly, transfer share measures the proportion of transfer passengers relative to total passengers, as expressed in Eq.~\eqref{eq:transferShare}. A greater O\&D local share indicates the majority of passengers flying on ATL $\rightarrow$ MEM route originate directly from ATL. Conversely, a smaller O\&D local share suggests that ATL is primarily utilized as a transfer hub, with a significant number of passengers connecting through ATL en route to MEM. This distinction is crucial for assessing the role of airports (e.g., ATL in this example) within the broader air transportation network, particularly in determining whether they serve primarily as points of origin or as intermediary hubs for passenger connections.
    
    \begin{equation}
    \label{eq:localShare}
    localShare_{{\text{ATL} \rightarrow \text{MEM}},t} = n_{local,t}/{n_{total,t}} = n_{1,t}/({n_{1,t}+n_{2,t}+n_{3,t}+n_{4,t}}).
    \end{equation}
    
    \begin{equation}
    \label{eq:transferShare}
    transferShare_{{\text{ATL} \rightarrow \text{MEM}},t} = n_{transfer,t}/{n_{total,t}} = (n_{2,t}+n_{3,t}+n_{4,t})/({n_{1,t}+n_{2,t}+n_{3,t}+n_{4,t}}).
    \end{equation}
    
    \subsection{Paper Organization}
    The rest of this paper is organized as follows: In Section~\ref{sec:lr}, related research work is reviewed, including research on local and transfer passengers, the impact and recovery from the pandemic on the air transportation industry, and clustering methods in related research. In Section~\ref{sec:Data}, data and data process will be introduced in details. In Section~\ref{sec:Analysis}, the patterns and dynamics of local share on the O\&D level will be analyzed in detail. Insights into local share characteristics are derived and discussed. In Section~\ref{sec:Methodology}, the clustering model development is fully discussed to further investigate the characteristics of local share on O\&Ds. In Section~\ref{sec:Conclusion}, the primary findings are summarized, and the implications of this study are discussed as well.

\section{Related Work Review}\label{sec:lr}
The related work for this research can be divided into two major categories. The first category focuses on the passenger demand analysis, specifically with regard to local and transfer passengers. The second category relates to the time series clustering methods used in the model development herein.

    \subsection{Local and Transfer Passenger Related Research}\label{sec:lr_business}
    % Literature of the related analysis
    % part 1 - the demand of analysis and forcast is the biggest part 
    % part 2 - there are some examples of local and transfer passenger literaures
    % Part 3 - some exasmples of the study after COVID 
    
    % part 1
    Tourism demand modeling and forecasting have been a significant area of interest for both researchers and practitioners over the past five decades~\cite{gunter2021forecasting}. Classic time series models, interpretable regression models, and advanced machine learning techniques have been extensively studied to identify significant changes in passenger demand at various levels of granularity and to forecast future passenger demand. As interest in more detailed analyses of passenger demand grows, particularly the composition of local and transfer passengers, a substantial body of research has emerged on this topic. Maertens et al. \citep{DevelopmentoTransferPassenger} developed a model to analyze transfer passenger data and transfer share globally, albeit at an airport-level (and not at the O\&D level). Their research revealed that different types of airports exhibit varying levels of transfer share. For example, major hubs located in less tourism-oriented cities, such as Atlanta and Doha, achieve higher transfer shares compared to gateway hubs in high-profile destinations like Dubai and New York. This empirical evidence underscores that local and transfer shares are specific to individual airports, with substantial heterogeneity in characteristics. While much of the literature focused on airport-level analyses of passenger share, such as the work by  Choi et al. \citep{DetermingFactorsofAirPassengers}, this research significantly extends the analysis by examining local and transfer share at the O\&D level. By focusing on higher granularity information, this study provides deeper insights into the distribution of local and transfer passengers and facilitates the integration of these findings into broader airport-level analyses.
    
    % part 2
    
    Local and transfer passenger shares are also crucial metrics for network planning and airline competition. The airport choice for transfer passengers is influenced by factors such as airfares, minimum connecting time (MCT), service quality of flight connection, and travel time. Attracting transfer passengers, in particular, is a primary objective in many airports' marketing strategies, as these passengers are not tied to a specific airport's local catchment area, and can easily switch among alternative connections offered by various airlines and hub airports. Consequently, airports face greater competition in the transfer passenger market than in the local O\&D market. Therefore, the development of the transfer passenger market is critical for airports, as transfer traffic supports the hub-and-spoke network, enabling airlines to sustain direct flights on routes that would not be feasible solely based on the local O\&D demand. Redondi et al. \citep{De-hubbingofAirportsandTheir} developed a model to identify de-hubbing (the process where a hub-and-spoke structure airline stops utilizing a certain airport as a connecting hub) airports and analyzed their recovery, studying hub characteristics through quantitative conditions by looking at the number of offered connections. Similarly, Wei and Hansen \citep{AnAggregateDemandModel} developed a model to predict demand in hub-and-spoke networks, considering airline service variables such as service frequency, aircraft size, ticket price, flight distance, the number of spokes in the network, influence of local passengers, as well as social-economic and demographic conditions in the metropolitan areas of the spokes and the hubs. Their study also identified the impact of local passengers on the hub-and-spoke network, demonstrating that airlines can attract more connecting passengers by increasing service frequency. Further, Zheng and Wei \citep{zheng2019air, zheng2020air} conducted data-driven analyses on direct passenger distribution in U.S. domestic O\&D markets, highlighting that the direct share is O\&D specific, for different O\&D pairs the pattern of direct passengers varies significantly. Their research utilized time series and classic machine learning models in forecasting direct passenger distribution, revealing that airline competition is a critical factor influencing direct passenger share.
    
    % part 3
    The significant impact of the COVID-19 pandemic in 2020 prompted extensive research into passenger demand recovery and forecasting. Tirtha et al. \citep{AnAirortLevelFramework} developed a model to examine the impact of COVID-19 on airline demand, incorporating factors in socio-demographic (population, education, age distribution), socio-economic (income, unemployment rate, GDP), built environment variables (number of trade centers, tourist attractions), level of service factors (average airfare and distance) and historical demand as lag variables. Gao \citep{gao2022benchmarking} employed airport-level passenger data to categorize airports into distinct groups according to their air travel demand recovery patterns during the pandemic. Their research indicated that air travel demand at most airports in the U.S. reached the lowest point in April 2020 and has gradually recovered since. While airports in southern regions have seen demand recover to pre-COVID levels by late 2021, many airports in regions such as the Eastern, the Great Lakes, and the New England regions have struggled to regain these levels.

    \subsection{Time Series Clustering Methods in Data Analysis}\label{sec:lr_model}

    Time series clustering is a technique used to uncover patterns and structures within sequential data~\cite{liao2005clustering}. Given the absence of established classification criteria for O\&Ds based on local share, a predefined ground truth for such classification is not readily available. As a result, we employ the following time series clustering to classify local share data by analyzing patterns in the time series. This method can reveal trends and abnormalities across different O\&D groups, offering valuable insights into demand forecasting and network management strategies~\cite{zanin2013modelling}. 

    Hierarchical clustering is a widely used technique in clustering analysis and has been adapted for time series data through the use of Dynamic Time Warping(DynTW)~\footnote{DynTW refers to Dynamic Time Warping algorithm, a time series analysis method, and should not be confused with the Detroit Metro Airport (DTW) airport code} as a distance metric. DynTW allows for flexibility in aligning time series that may exhibit temporal misalignments~\cite{berndt1994using}. In aviation research, DynTW has been applied to cluster flight trajectories, helping to identify similar patterns and commonalities in aircraft movements~\cite{todoric2023comparison, Zhang2024}. Moreover, Wang et al. \citep{wang2023similarity} used DynTW to recognize pilots' operation process sequences during flight tasks to detect operational errors and improvements in overall safety measures.
    
    The $k$-shape clustering algorithm, proposed by Paparrizos and Gravano  \citep{paparrizos2015k}, is a generalization of the $k$-means algorithm that can handle time series data. Unlike traditional $k$-means clustering, the $k$-shape algorithm uses a normalized cross-correlation measure as the distance metric, which also addresses the issue of temporal misalignments found in time series data. For instance, Gao \citep{gao2022benchmarking} applied the $k$-shape algorithm to classify airports based on patterns of air travel demand recovery during the COVID-19 pandemic. The algorithm is effective in identifying clusters of airports that exhibited similar recovery trends, providing valuable insights into post-pandemic recovery strategies.

    Additionally, Self-Organizing Maps (SOMs) are a type of neural network designed to reduce the dimensionality of data while performing clustering based on the topological properties of the input space~\cite{kohonen1990self}. In SOMs, input data vectors are mapped to the closest neuron in a grid-like structure, and similar data points tend to be mapped to nearby neurons, creating a two-dimensional topological map that can reveal inherent patterns within time series data. Kumar \citep{kumar2014self} applied SOMs to airspace sectorization, demonstrating how this technique could improve air traffic management by creating balanced airspace sectors, thereby reducing collision risks and optimizing air traffic flow.

    Shape-based distance (SBD) is another measure used to calculate the similarity between two time series by employing a normalized version of cross-correlation that is invariant to scaling and translation. This indicates that it can effectively handle differences in magnitude or time shifts between series~\cite{paparrizos2017fast}. When combined with Affinity Propagation (AP), an algorithm that automatically determines the optimal number of clusters based on the input data, SBD proves to be effective in clustering time series without the need to specify the number of clusters beforehand~\cite{frey2007clustering}. For example, El-Samak and Ashour \citep{el2015optimization} applied AP clustering to optimize the traveling salesman problem, identifying cities that are efficiently accessed by travelers.

    Furthermore, DynTW Barycenter Averaging (DBA) combined with Gaussian Mixture Models (GMMs) provides an approach for soft clustering of time series data~\cite{petitjean2011global}. In soft clustering, GMMs assign probabilities to each data point's membership in a cluster, rather than a hard, definitive classification. This method is especially beneficial when the time series data exhibits overlap across clusters. 

    As clustering is an unsupervised learning method with no predefined ground truth, it is essential to evaluate the clustering performance using various metrics to ensure the quality and validity of the results. The following evaluation metrics are commonly employed in clustering evaluations:

    \begin{itemize}
        \item \textbf{Silhouette Score:} This metric measures how similar a data point is to its own cluster compared to other clusters~\cite{rousseeuw1987silhouettes}. It ranges from -1 to 1, with a higher score indicating better clustering performance. A higher score suggests that data points are well-matched to their clusters and poorly matched to neighboring clusters.
        \item \textbf{Davies-Bouldin Index:} This index quantifies the average similarity between each cluster and its most similar cluster, focusing on the worst-case pair of clusters~\cite{davies1979cluster}. A lower score is desirable, as it indicates more distinct and well-separated clusters.
        \item \textbf{Dunn Index:} This index evaluates the ratio of the minimum inter-cluster distance to the maximum intra-cluster distance~\cite{dunn1974well}. A higher score reflects better clustering, emphasizing compact clusters that are far apart from each other.
        \item \textbf{Calinski-Harabasz Index:} This index compares the overall inter-cluster variance to the intra-cluster variance~\cite{calinski1974dendrite}. A higher score indicates better clustering performance, as it suggests a higher degree of separation between clusters relative to the internal cohesion within clusters.
    \end{itemize}

    \subsection{Research Gaps}

    Existing research in air transportation passenger flow analysis has predominantly emphasized aggregate demand forecasts, airline competition, and macroeconomic factors, with limited differentiation among passenger categories. Moreover, the current air passenger demand studies often focus on broad metrics at the airport- or system-level, failing to capture the nuanced variations that occur at the O\&D pair level. In-depth research on local and transfer passengers on the O\&D level is becoming increasingly crucial for airline marketing and network planning, airport investment, and authority decision making. A comprehensive understanding of the distinctions between local and transfer passengers remains underdeveloped, leaving a critical gap in the ability to analyze passenger composition and its evolution over time. To address this need, this study examined local share on O\&D level, analyzing the factors driving changes in the local and transfer passenger demand change.

\section{Data and O\&D Pair Selection}\label{sec:Data}
    \subsection{Data} \label{sec:Data_DB1BT100}
    This research utilized publicly available datasets from the Bureau of Transportation Statics (BTS), specifically the Airline Origin and Destination Survey (DB1B) and Air Carrier Statistics (T100). These datasets were employed to create the local share metric and other related variables. 

    The DB1B database represents a 10\% sample of quarterly airline ticket data reported by U.S. air carriers~\cite{DB1B}. This dataset is organized into three primary tables: D1B1Coupon, DB1BTicket, and DB1BMarket. The DB1BCoupon table contains coupon-specific information for each domestic itinerary. The DB1BTicket table records round-trip level data. The DB1BMarket table used in this research, captures data at the directional market level, providing information on single trips within an O\&D market. It includes details such as the reporting carrier, origin and destination airports, number of market coupons, and market miles flown. The connection information on an itinerary is one of the key information from the DB1BMarket data table used in this research. 

    T100 data is another critical dataset, providing comprehensive information on U.S. domestic airline operations. It includes details on both flight segments and flight markets, with variables such as airline, origin, destination, passenger counts, departures, seats, and other operational data. T100 is essential for analyzing domestic travel patterns, airline performance, and market trends, aiding in transportation planning and policy development~\cite{teixeira2020revealing, li2019recent, bardell2023quantifying}. In this research, the T100 dataset was utilized flexibly to generate a variety of air transportation metrics, including daily departure, direct flight services, and fleet mix.

    \subsection{O\&D Pair Selection}
    The DB1BMarket database contains comprehensive information reported by all carriers, capturing a wide range of domestic markets. The initially developed O\&D local share dataset includes all domestic markets reported, resulting in a vast number of directional O\&D pairs. For instance, in the first quarter of 2023, 437 domestic airports were reported, accounting for 64,173 directional O\&D pairs. However, a considerable number of these directional O\&D pairs are not consistently active, necessitating a careful selection process to ensure the robustness of the analysis: Firstly, the FAA airport categorization~\cite{AirportCategoriesAirports} is considered, which classifies domestic airports into large-, medium-, small-, and non-hubs. O\&D pairs connecting at least one large-hub airport are prioritized for inclusion. Examples of large-hub airports include ATL, ORD (Chicago O'Hare International Airport, Chicago, IL), and DFW (Dallas Fort Worth International Airport, Dallas, TX). Secondly, only directional O\&D pairs that have maintained local passengers consistently (local share is greater than 0) since 2021 were selected. This criterion ensures that long-inactive O\&D pairs, which had ceased operations or lacked sufficient local passenger flow, were excluded from the analysis. By focusing on consistently active O\&D pairs, the refined dataset represents the most relevant and operationally significant routes in the domestic air transportation system.

    Applying these criteria resulted in the identification of domestic 3,936 O\&D pairs, accounting for 90\% of domestic passenger traffic between Jan. 1st, 2021 and Dec. 31st, 2023. Although this process significantly reduced the number of O\&D pairs, the selected O\&D pairs still covered the majority of air passenger traffic across the system, ensuring that the analysis remained comprehensive while focusing on the important O\&D pairs.

\section{Local Share Analysis}\label{sec:Analysis}

To comprehensively investigate the dynamics of the local share, this study undertook a series of analyses of O\&D local share time series with different characteristics. Additionally, correlating factors were identified to elucidate the temporal changes in local share, offering insights into how passenger composition evolves over time. 

    \subsection{Seasonal Dynamics and Hub Utilization}\label{sec:Analysis_MSPANC}
    
\begin{figure}[htbp]
        \centering
        % Subfigure 1
        \begin{subfigure}[b]{0.45\textwidth}
            \centering
            \includegraphics[width=\textwidth]{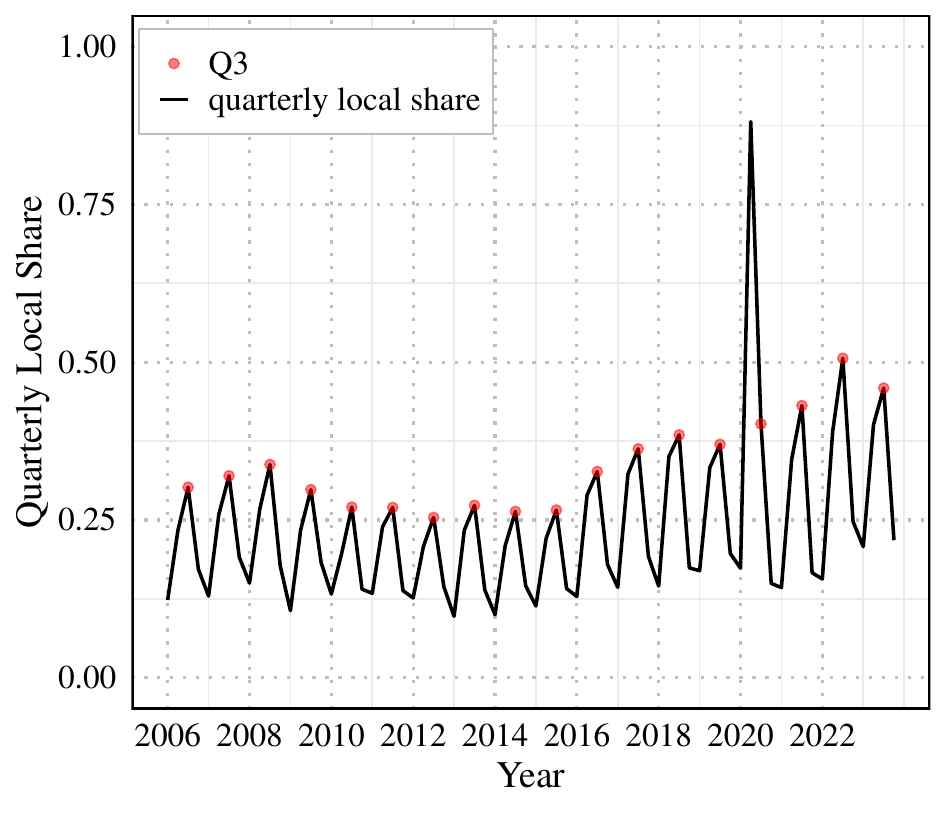}
            \caption{Local share time series.}
            \label{fig:LocalShare_MSPANC}
        \end{subfigure}
        ~
        % Subfigure 2
        \begin{subfigure}[b]{0.45\textwidth}
            \centering
            \includegraphics[width=\textwidth]{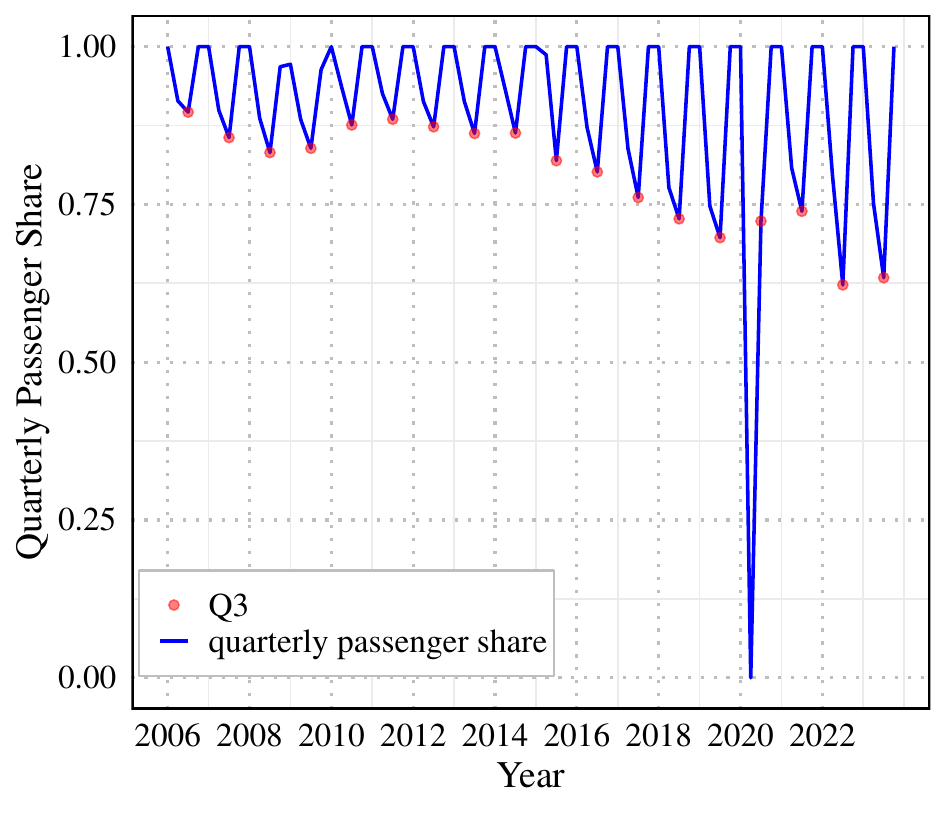}
            \caption{Northwest and Delta passenger share time series.}
            \label{fig:PaxShare_MSPANC}
        \end{subfigure}
        % \vspace{0.5em} 
        \caption{Local share and passenger share on MSP $\rightarrow$ ANC.}
        \label{fig:Example_MSPANC}
    \end{figure}

    ANC is the largest and busiest airport in Alaska. It handles the highest volume of passenger traffic in the state and serves as a critical hub for international cargo operations due to its strategic location between North America and Asia. It accommodates a variety of carriers, including Alaska Airlines (Alaska), Delta Air Lines (Delta), and numerous international airlines. It also experiences pronounced increases in passenger activity during the summer tourist season~\cite{yu2009effects}. Direct flight services to ANC originating from a limited number of U.S. airports are served by certain carriers, such as DEN (Denver International Airport, Denver, CO) $\rightarrow$ ANC by United Airlines (United), MSP (Minneapolis–Saint Paul International Airport, St Paul, MN) $\rightarrow$ ANC by Delta, and PHX (Phoenix Sky Harbor International Airport, Phoenix, AZ) $\rightarrow$ ANC by Alaska. Figure~\ref{fig:LocalShare_MSPANC} presents the local share time series on MSP $\rightarrow$ ANC. MSP, situated in the Twin Cities region of Minnesota is one of the largest airports in the Midwest and a key hub for Delta Air Lines. MSP offers numerous domestic and international flights and features operations as a significant connecting point for flights within the U.S. and to Canada.
    
    % why there is a strong seasonality on this OD
    There is a strong pattern of seasonality in the local share on MSP $\rightarrow$ ANC. As Anchorage is a popular summer destination, passenger volumes increase significantly during warmer months. Prior to its merger with Delta in 2009, Northwest Airlines (Northwest) was the primary provider of both direct and transfer flights through MSP to ANC. After the merger, Delta assumed this role. Figure~\ref{fig:PaxShare_MSPANC} shows the passenger share of Northwest and Delta on MSP $\rightarrow$ ANC over time, highlighting that their market passenger share is lower in the second and third quarters. During the summer, additional regional carriers, such as Sun Country Airlines, offer direct services from MSP to ANC, effectively increasing the direct service capacity. Through an analysis of direct flights using T100 data, we observed that Northwest Airlines, and later Delta, consistently increase direct flights from MSP to ANC during the summer. This seasonal adjustment caters to the influx of international passengers connecting through MSP en route to Anchorage, ensuring smoother travel for these travelers.
    
    % a small piece of summary about this type of example 
    Markets such as MSP $\rightarrow$ ANC, which connect destinations with strong seasonal appeal, exhibit pronounced seasonal variations in both local and transfer passenger segments. During peak travel seasons, these markets attract increased competition, with LCCs, ULCCs, and regional carriers entering to provide direct flight services. Competitive airfares and shorter travel times, often facilitated by geographical advantages, encourage passengers to select direct flight services, thereby increasing the local share in certain markets. There are about 150 domestic airports connecting to ANC through MSP, and Delta effectively utilized MSP as a central point within its network. The passengers channeling through MSP to reach ANC, which contributes to a local share on MSP $\rightarrow$ ANC fluctuating around 0.2. This pattern vividly illustrates how legacy carriers leveraged their hubs to efficiently distribute passengers across their route networks.
    
    \subsection{Regional Competition and Airline Strategy Shifts}\label{sec:Analysis_BWIALB}
\begin{figure}[htbp]
        \centering
        % Subfigure 1
        \begin{subfigure}[b]{0.45\textwidth}
            \centering
            \includegraphics[width=\textwidth]{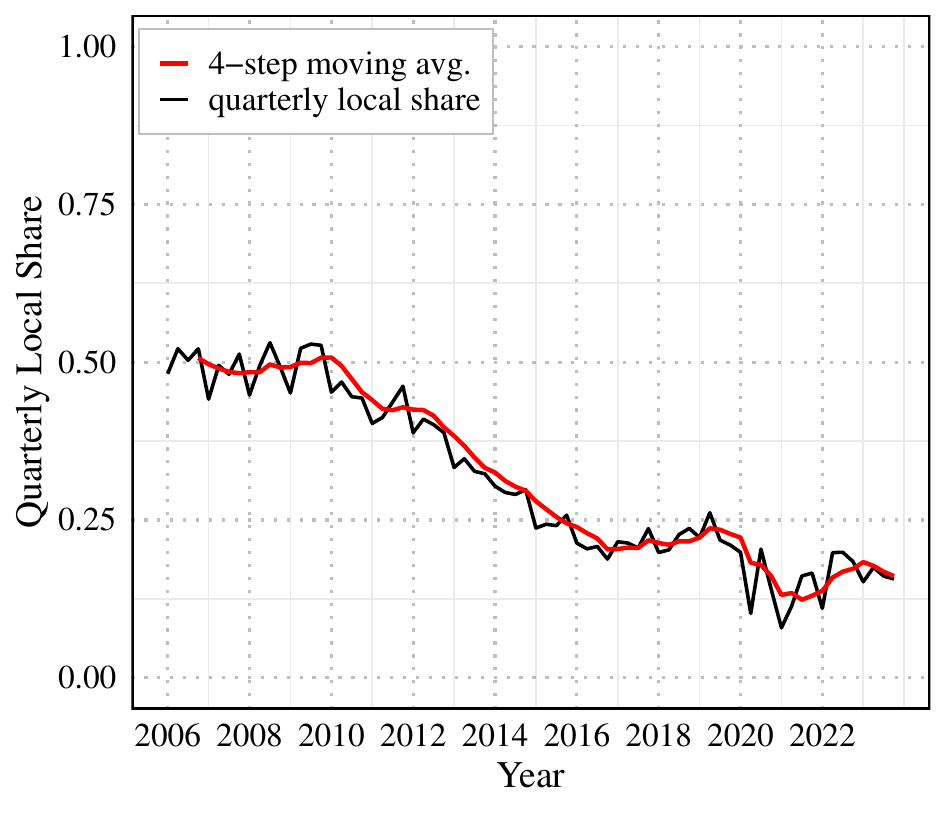}
            \caption{Local share time series.}
            \label{fig:LocalShare_BWIALB}
        \end{subfigure}
        ~
        % Subfigure 2
        \begin{subfigure}[b]{0.45\textwidth}
            \centering
            \includegraphics[width=\textwidth]{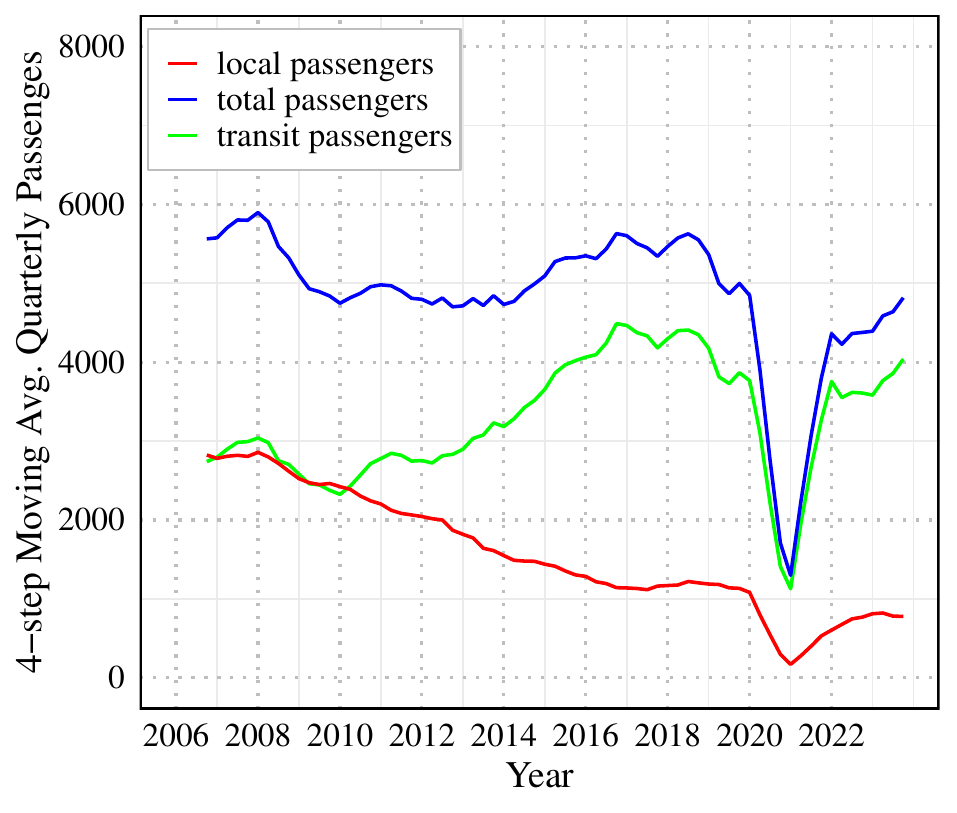}
            \caption{Passenger in different categories time series.}
            \label{fig:PaxChange_BWIALB}
        \end{subfigure}
        % \vspace{0.5em} 
        \caption{Local share and passenger change BWI $\rightarrow$ ALB.}
        \label{fig:Example_BWIALB}
    \end{figure}

    BWI (Baltimore/Washington International Thurgood Marshall Airport, Baltimore, MD) $\rightarrow$ ALB (Albany International Airport, Albany, NY) presents another illustrative example. Figure~\ref{fig:LocalShare_BWIALB} shows the quarterly local share time series on BWI $\rightarrow$ ALB. BWI serves as a major airport in the Washington, D.C. metropolitan area and functions as a principal hub for Southwest Airlines (Southwest). ALB, located in the Capital Region of Northeastern New York and Western New England, represents a market served exclusively by Southwest Airlines. The local share steadily declined from 0.51 in 2009 to 0.20 in 2016, and with the impact of the COVID-19 pandemic, it further decreased to 0.15 by 2023. 
    
    The decline in local share on BWI $\rightarrow$ ALB reflects a complex interplay of market forces, competitive strategies, and evolving passenger behaviors. As illustrated in Figure~\ref{fig:PaxChange_BWIALB}, the number of local and transfer passengers remained relatively balanced until 2010. However, from 2010 to 2016, the local share steadily declined despite stable total passenger volumes. During this period, local passengers decreased by 54\%, while transfer passengers increased by 47\%. This shift in passenger composition coincided with intensified regional competition and strategic changes by Southwest. There are two other major airports in the Washington, D.C. metropolitan area, which are DCA and IAD (Washington Dulles International Airport, Dulles, VA). Both DCA and IAD offer direct flights to ALB through US Airways, and later after merge, American Airlines (American) at DCA and United at IAD. Between 2010 and 2016, local passenger volumes from these two airports to ALB were more than doubled. Simultaneously, Southwest Airlines enhanced BWI's role as a transfer hub, increasing the number of origin airports offering service to ALB via BWI from 51 in 2009 to 60 in 2016.

    This case study underscores the multifaceted nature of local share dynamics. As airlines adjust their networks, to capitalize on market opportunities and respond to competitive pressures, passenger decisions regarding which airport to use and whether to select direct or connecting services are shaped by the interplay of price, travel time, and available connections. In turn, the evolution of local share serves as an indicator of how airline network strategies, market competition, and passenger preferences interact within the air transportation system.

    \subsection{O\&D with Disparate Local Share}\label{sec:Analysis_PBITPA}

\begin{figure}[htbp]
        \centering
        % Subfigure 1
        \begin{subfigure}[b]{0.45\textwidth}
            \centering
            \includegraphics[width=\textwidth]{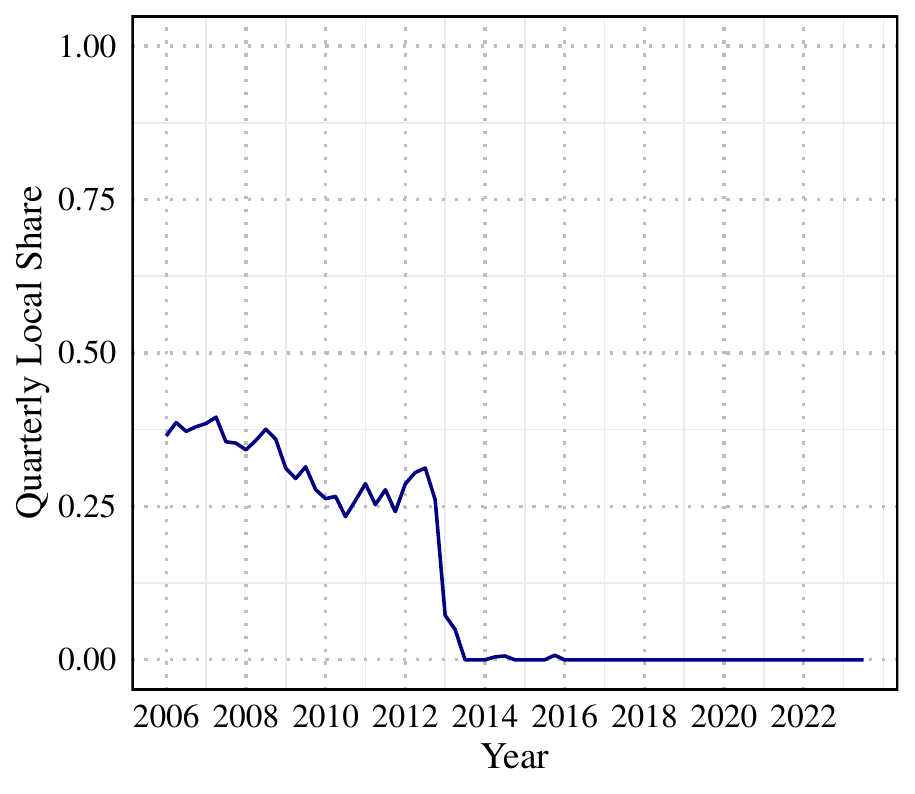}
            \caption{Local share time series on TPA $\rightarrow$ PBI.}
            \label{fig:LocalShare_TPAPBI}
        \end{subfigure}
        ~
        % Subfigure 2
        \begin{subfigure}[b]{0.45\textwidth}
            \centering
            \includegraphics[width=\textwidth]{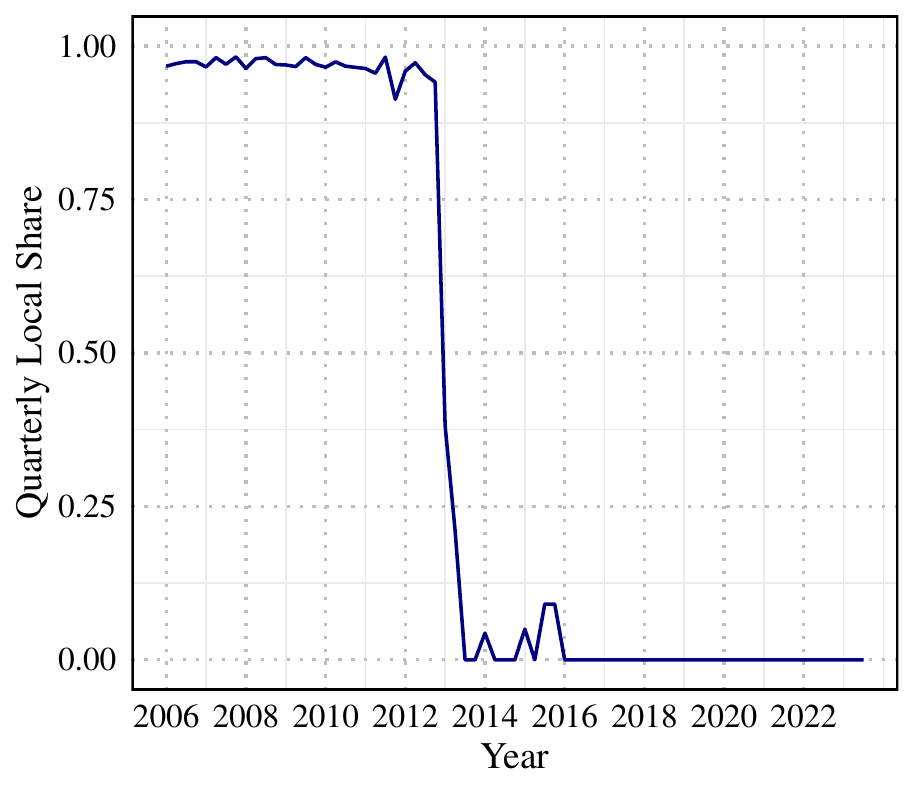}
            \caption{Local share time series on PBI $\rightarrow$ TPA.}
            \label{fig:LocalShare_PBITPA}
        \end{subfigure}
        % \vspace{0.5em} 
        \caption{Local share on O\&D pairs PBI $\longleftrightarrow$ TPA.}
        \label{fig:Example_PBITAP}
    \end{figure}

    A more pronounced example of a decline in local share is observed in Figures~\ref{fig:LocalShare_TPAPBI} and Figure~\ref{fig:LocalShare_PBITPA}, which depict the local share time series on TPA (Tampa International Airport, Tampa, FL) $\rightarrow$ PBI (Palm Beach International Airport, West Palm Beach, FL) and PBI $\rightarrow$ TPA, respectively. In both cases, the local share dropped to 0 in 2013. Southwest had predominantly served these O\&D pairs, with Silver Airways providing a comparatively smaller portion of flights concurrently. However, beginning in 2013, Southwest withdrew entirely from this market. The limited number of passengers on this O\&D, as reflected in the DB1B sample, further emphasizes the sensitivity of local shares to changes in airline operations. In contrast, the O\&D route from PBI to TPA maintains a significantly higher local share than the reverse direction from TPA to PBI prior to Southwest's exit. This disparity can be attributed to the differing roles of the two airports and the nature of their surrounding regions. Tampa, a prominent cruise embarkation hub, draws a significant number of tourists. This robust demand encourages airlines to use TPA as a hub for directing traffic to other airports across Florida. On the other hand, West Palm Beach, catering to affluent seasonal residential communities, does not support the same scale of large-scale tourist attractions. Consequently, as carriers revised their operations and the market structure evolved over time, the local share adjustments at both ends of the route underscored how airport characteristics, destination appeal, and airline strategies collectively shaped passenger composition within O\&D markets.

    \subsection{Network Reconfiguration and Carrier Realignment}\label{sec:Analysis_MEMDTW}
\begin{figure}[htbp]
        \centering
        % Subfigure 1
        \begin{subfigure}[b]{0.45\textwidth}
            \centering
            \includegraphics[width=\textwidth]{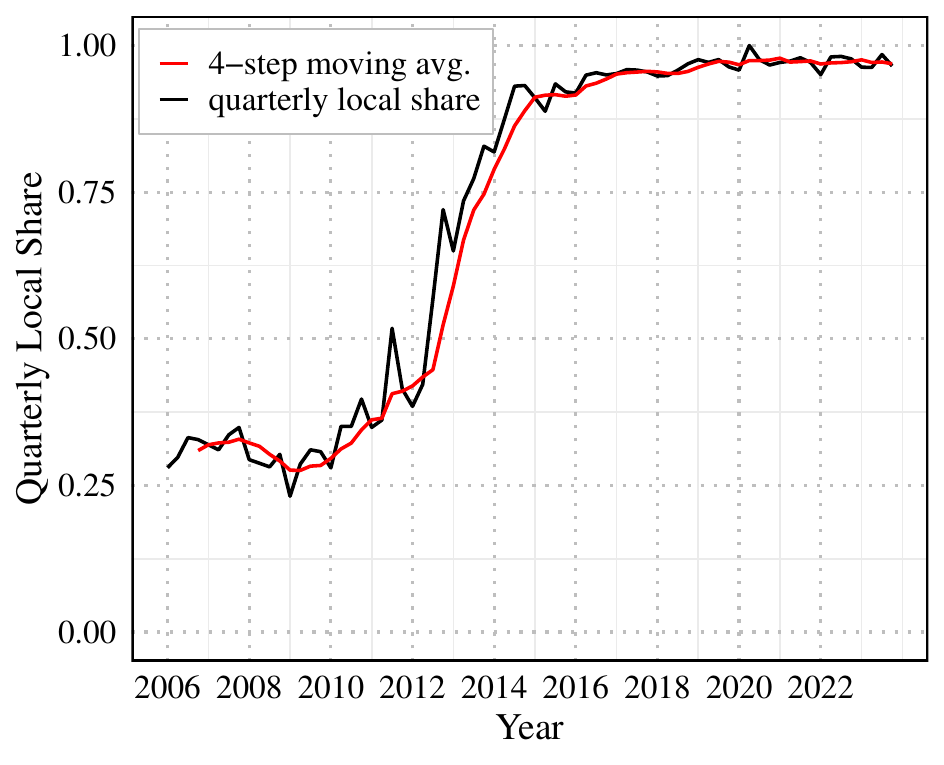}
            \caption{Local share time series.}
            \label{fig:LocalShare_MEMDDTW}
        \end{subfigure}
        ~
        % Subfigure 2
        \begin{subfigure}[b]{0.45\textwidth}
            \centering
            \includegraphics[width=\textwidth]{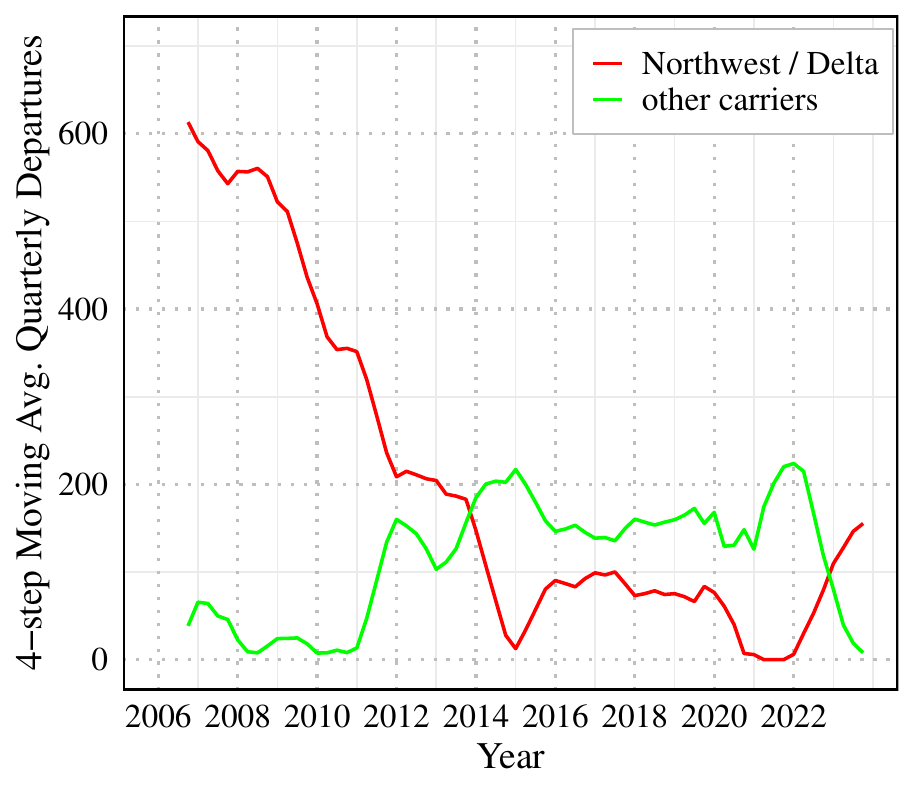}
            \caption{Actual departures time series by different carriers.}
            \label{fig:Departure_MEMDTW}
        \end{subfigure}
        % \vspace{0.5em} 
        \caption{Local share and actual departure on MEM $\rightarrow$ DTW.}
        \label{fig:Example_MEMDTW}
    \end{figure}

    An additional example illustrating significant fluctuations in local share over time is MEM $\rightarrow$ DTW (Detroit Metropolitan Wayne County Airport, Detroit, MI). MEM is one of the world's largest cargo airports and the primary hub of FedEx Express. It offers a variety of domestic and international flights, operated by legacy carriers such as Delta Air Lines and American Airlines. DTW is among the busiest airports in the United States, and is classified as a medium-sized hub. It functioned as a major hub for Delta, and supports numerous domestic and international services. Figure~\ref{fig:LocalShare_MEMDDTW} depicts the local share time series on MEM $\rightarrow$ DTW over time. 

    The local share on MEM $\rightarrow$ DTW passed through three distinct phases. Initially, it stabilized at approximately 0.3 until 2012. Between 2012 and 2014, the local share value rose sharply to about 0.95 within these two years, and the elevated level persisted even throughout the pandemic. The dramatic increase indicated a fundamental shift in the market dynamics, prompting a closer examination to identify the factors responsible for this transformation. Following the merger with Northwest, Delta implemented a strategic realignment, focusing on network optimization, hub realignment, and market demand, which included reducing its operations at MEM \citep{DeltaMerge}. This downsizing reshaped MEM from a major Delta hub to a more regionally oriented airport with fewer routes. Figure~\ref{fig:Departure_MEMDTW} presents the quarterly departure counts by different carriers, highlighting the reduction in departures from MEM to DTW in 2012 through 2014. As Delta decreased operations, regional carriers such as GoJet Airlines, Endeavor Air, and SkyWest Airlines introduced service on MEM $\rightarrow$ DTW. Notably, while most regional carriers later withdrew, Delta continued to serve the route post-pandemic, illustrating how the carrier's strategic shifts influenced the competitive environment over time.

\begin{figure}[hbt!]
        \centering
        % Subfigure 1
        % \begin{subfigure}[b]{0.8\textwidth}
        \begin{subfigure}[b]{0.7\textwidth}
            \centering
            \includegraphics[width=\textwidth]{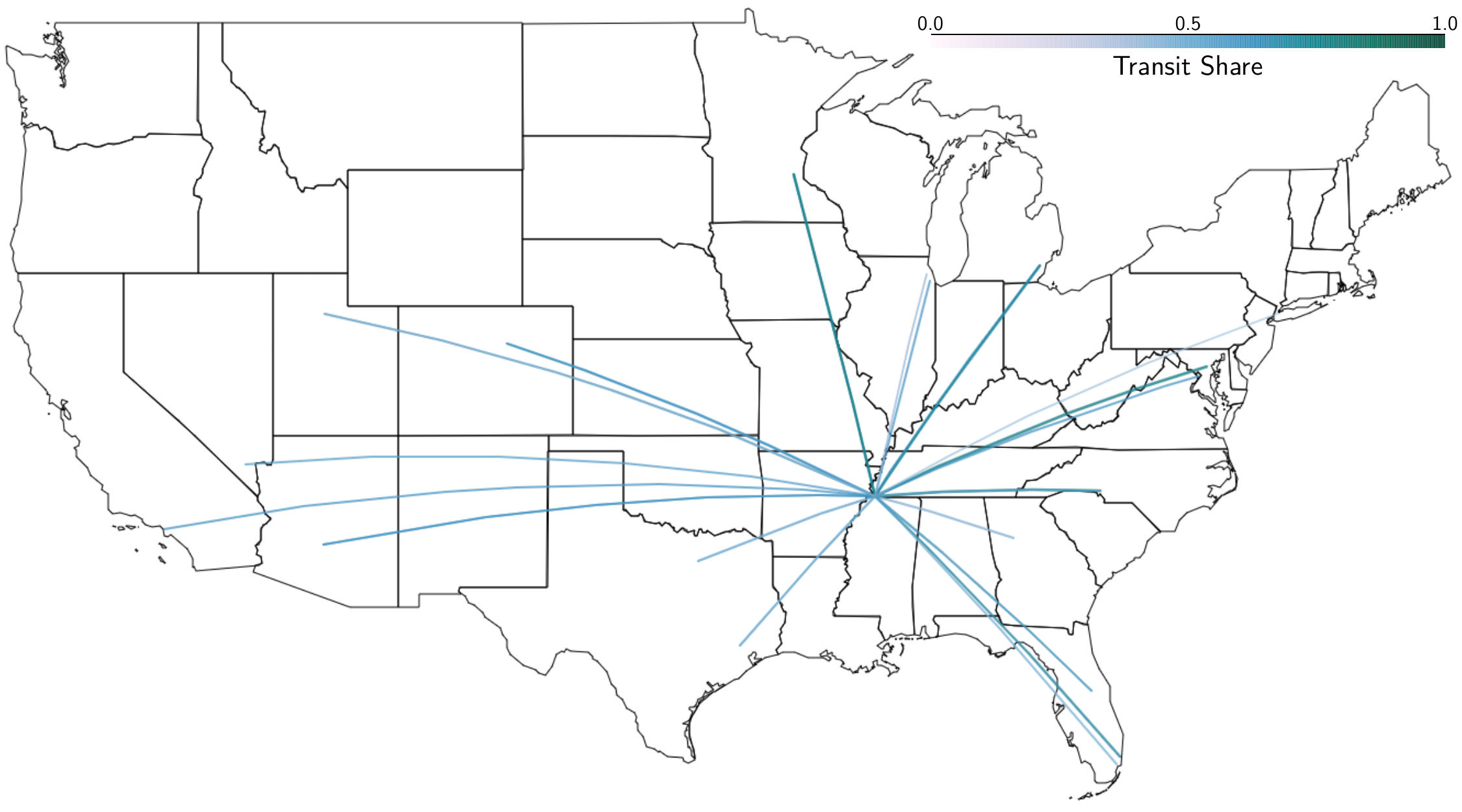}
            \caption{2006.}
            \label{fig:LS_MEMMarkets_2006}
        \end{subfigure}
        % ~
        % Subfigure 2
        % \begin{subfigure}[b]{0.8\textwidth}
        \begin{subfigure}[b]{0.7\textwidth}
            \centering
            \includegraphics[width=\textwidth]{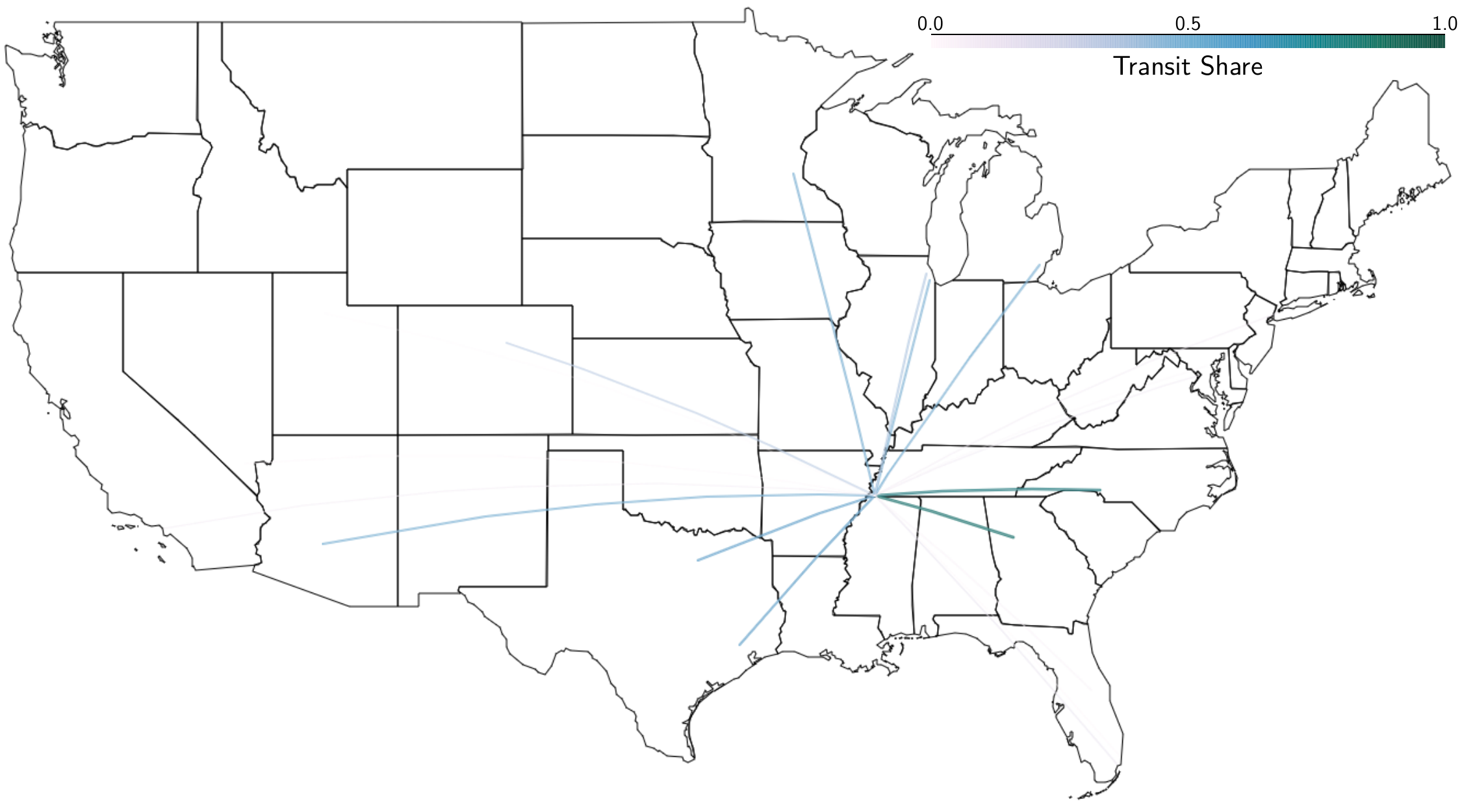}
            \caption{2008.}
            \label{fig:LS_MEMMarkets_2018}
        \end{subfigure}
        % \vspace{0.5em} 
        \caption{Direct connections with MEM and transfer share.}
        \label{fig:LS_MEMMarkets}
    \end{figure}

    % further descussion about the local hare change 
    The impact of Delta Air Lines' strategic withdrawal at MEM was not confined to the MEM $\rightarrow$ DTW route. Figure~\ref{fig:LS_MEMMarkets} depicts transfer share trends for various O\&D markets involving MEM. Comparing 2006 to 2018, a substantial reduction in direct O\&D connections occurred, especially to markets such as Los Angeles on the West Coast, major airports in Florida, and East Coast gateways like DCA. Among the remaining O\&D markets, transfer share declined notably, including routes such as MEM $\rightarrow$ DEN, and MEM $\rightarrow$ PHX. 
    
    This example underscores that changes in airline strategy could profoundly alter local and transfer passenger dynamics across multiple O\&D pairs. When a legacy airline reduces its footprint, it can create openings for LCCs, ULCCs, and regional carriers specializing in P2P services. Conversely, a legacy airline's increased involvement could have widespread effects, reshaping travel patterns throughout the connected network.

    From the example discussed above, the O\&D local share displays the interplay between seasonal market characteristics, airport and airline competition, hub utilization, de-hub effects, and passenger preferences, which makes the local share a valuable indicator of evolving market conditions and industry strategies.

\section{Methodology}\label{sec:Methodology}

In this research, local share analysis encompassed 3,936 O\&D pairs across the U.S. domestic air transportation system. To systematically group O\&Ds exhibiting similar local share characteristics and efficiently identify patterns in local share time series, various time series clustering techniques were explored. 

    \subsection{Problem Modeling}
    Rather than examining seasonal fluctuations in the local share time series, we focus on the long-term patterns and characteristics observed over the local share time series. The quarterly O\&D local share time series from Q1 2006 to Q3 2024 were used, and the data were aggregated by year to minimize the impact of seasonality. This aggregation better reflects long-term annual trends in local share. In this part, the primary interest lies in identifying underlying annual patterns rather than short-term seasonal fluctuations. By focusing on aggregated yearly data, the analysis emphasized the trend and shape of the local share time series for each O\&D pair, enabling the identification of O\&D groups with similar market behaviors. The local share metrics were recalculated based on the yearly sum of local passengers divided by the yearly sum of total passengers. Other attributes such as the airport type were also retained for further analysis. These attributes were considered potentially influential factors in the clustering analysis and subsequent interpretation of the results.
    
    Various unsupervised clustering methods, as discussed in Section \ref{sec:lr_model}, were applied to the dataset to identify the characteristics of O\&D pairs based on the local share time series. The temporal data were either standardized or left un-standardized, depending on the clustering method used. Standardization was performed across the full time series for each O\&D pair considering all years, rather than across all O\&D pairs within a given year. This approach was chosen to ensure that the clustering results for standardized data were based solely on the local share trend of each O\&D pair, rather than the magnitude of the local share. 

    \subsection{Model Development}
        \subsubsection{Hierarchical Clustering with DynTW}
        
        Hierarchical clustering began with the computation of pairwise DynTW distances between all O\&D pairs. The resulting pairwise distance matrix can be organized into a condensed form via a 1D array that stores only the upper triangular values of the pairwise distance matrix, excluding redundant entries. This condensed form of matrix was normalized by dividing the maximum distance to ensure that the distance values were on a comparable scale. Any single large-distance value could significantly affect the stability of the clustering results. 
        
        Ward's linkage method was selected to build the hierarchical clustering tree, as it minimizes the total within-cluster variance, leading to more compact and homogeneous clusters~\cite{murtagh2012algorithms}. Clusters were formed by cutting the hierarchical tree at a specific height, which was determined by the number of clusters desired. Different numbers of clusters were specified to explore the sensitivity of the clustering results at different levels of granularity. However, it is worth noting that such a height-based approach may not always yield the desired number of clusters. The occurrence of remaining empty-alike clusters was due to the hierarchical clustering's agglomerative nature and the inherent distribution of the data. Specifically, in cases where a small, isolated cluster (such as a single outlier) would cause a large increase in variance when merged with a larger cluster, the Ward algorithm postpones or avoids this merge~\cite{müllner2011modern}. As a result, small or isolated clusters may contain few data points that are either dissimilar to others or too distant to merge without violating the method's variance criteria.

        \subsubsection{$k$-Shape Clustering}
        
        $k$-Shape clustering was applied to capture shape-based similarities regardless of amplitude differences in the local share time series~\cite{paparrizos2015k}. This characteristic makes $k$-Shape clustering suitable for time series data where the shape of the data conveys more information than the magnitude. For example, there exist two temperature time series from different cities, one with consistent daily fluctuations and the other with large seasonal variations but similar daily trends. Despite differing temperature ranges (amplitude), the time series may share similar daily patterns. $k$-Shape clustering can group these series based on shape, disregarding magnitude differences. The algorithm iteratively refined the cluster assignments by maximizing the shape-based similarity between the time series and their respective cluster centroids. Therefore, if the data do not exhibit clear shape-based patterns, the algorithm may also converge to a solution with fewer effective clusters, leaving some clusters empty.

        \subsubsection{Self-Organizing Maps}
        
        The Self-Organizing Maps (SOM) algorithm was configured with grid dimensions of $4 \times 4$ and $10 \times 10$ (i.e., 16 neurons and 100 neurons) to explore different levels of granularity. The choice of grid size affects the number of clusters, with larger grids allowing for finer distinctions between patterns. It is also important to note that the choice of grid dimensions was guided by exploratory analysis but may still not be perfectly aligned with the intrinsic data patterns. The SOM algorithm iteratively adjusted the weights of the neurons over 1,000 epochs with a learning rate of 0.5 and a neighborhood function defined by the Gaussian function. Each O\&D pair was assigned to the neuron with the closest weight vector, and the distribution of O\&D pairs across the SOM grid was analyzed to interpret the clustering outcomes.

        \subsubsection{Shape-based Distance with Affinity Propagation}

        The Shape-based Distance is a measure specifically designed for time series data, capturing shape-based similarities while being robust to amplitude and offset differences. Although $k$-Shape also relies on SBD, it differs in how clusters are formed: $k$-Shape is centroid-based and requires specifying the number of clusters in advance, whereas Affinity Propagation (AP) identifies \emph{exemplars}—representative data points—without a predefined number of clusters~\cite{frey2007clustering}. To apply AP, we convert the SBD values into a similarity matrix by taking the negative of the distance (i.e., the lower distance becomes the higher similarity). AP automatically discovered 190 clusters from the dataset. Additionally, to further condense the results into fewer interpretable groups, a second clustering step using Hierarchical Clustering (Average linkage~\cite{yang201719Temporal}) was performed on the exemplars. This two-step approach tried to leverage AP's ability to find fine-grained clusters first, and then merged closely related exemplars into broader clusters, providing a clearer overview of the data's shape-based patterns.

        \subsubsection{DynTW Barycenter Averaging with Gaussian Mixture Models}
        
        The combination of DynTW Barycenter Averaging (DBA) with Gaussian Mixture Models (GMM) is a two-step process to handle both time series alignment and probabilistic clustering. First, DBA computes an average sequence (barycenter) that minimizes the sum of DynTW distances to all input sequences within each cluster~\cite{petitjean2011global}, thus capturing their main patterns despite temporal misalignments. Next, a time series $k$-means algorithm refined these barycenters and cluster assignments by minimizing inertia, i.e., how tightly each time series is grouped around its barycenter. These initial labels tried to address the GMM's sensitivity to the Gaussianity assumption and cluster initialization~\cite{reynolds2009gaussian}. The GMM is then configured with the same target number of clusters (e.g., $k=10$), by incorporating the initial cluster labels via one-hot encoding to identify both local and global patterns in the local share time series data.

    \subsection{Model Analysis}

    The clustering results obtained from the various unsupervised methods involved a combination of qualitative visualizations and quantitative performance metrics, aiming to systematically evaluate their effectiveness and interpretability. Challenges and limitations encountered during the clustering process were also discussed to provide insights into the practical application of the methods. Additionally, a closer analysis of individual cases were conducted to gain deeper insights into specific O\&D pairs within the clusters.

        \subsubsection{Clustering Visualization Overview with Individual Case Analysis}
\begin{figure}[htbp]
                \centering
                % Subfigure 1
                \begin{subfigure}[b]{0.47\textwidth}
                    \centering
                    \includegraphics[width=\textwidth, trim={0cm 0.2cm 0cm 1.2cm}]{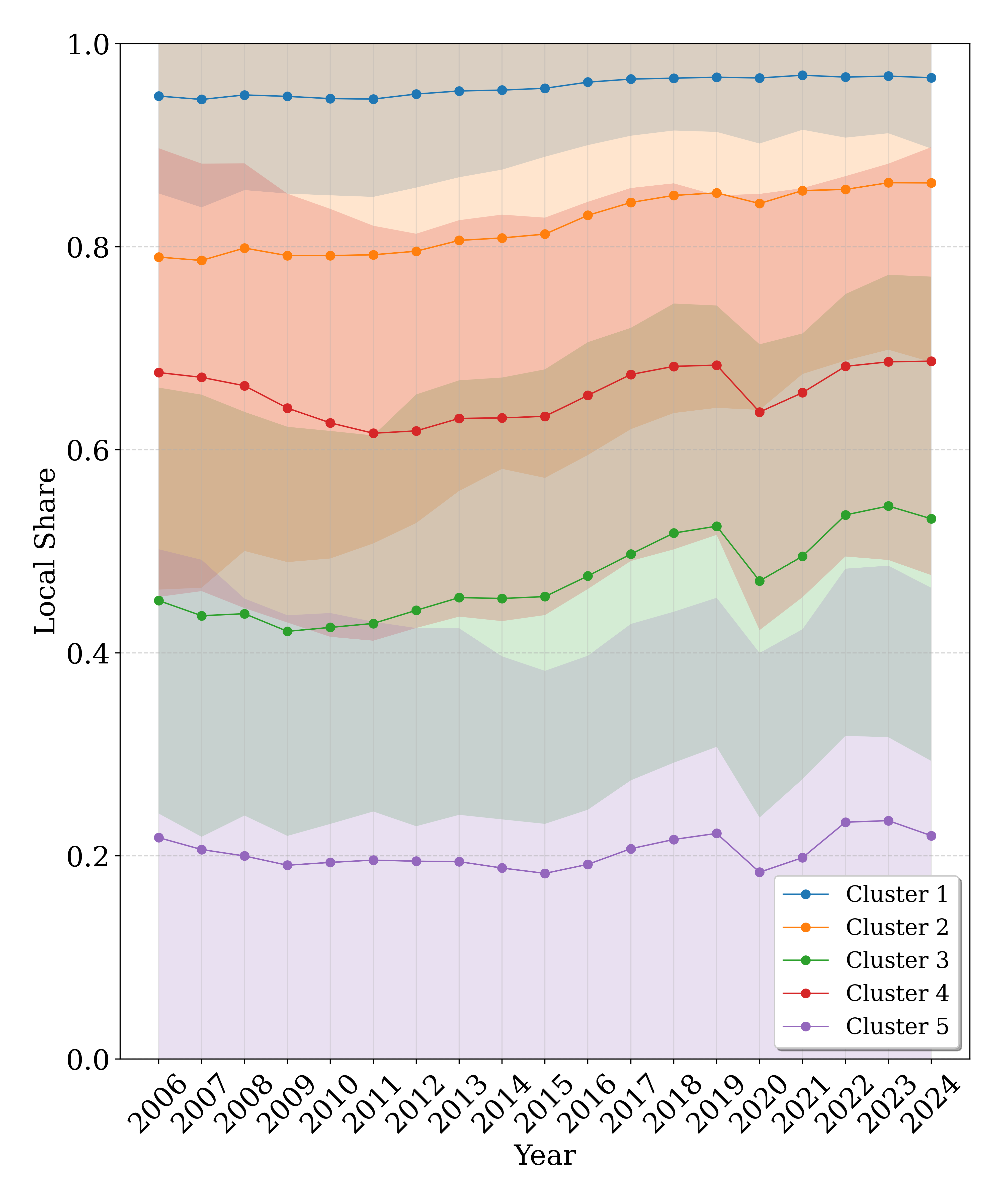}
                    \caption{Average values with 95\% conf. inv.}
                    \label{fig:dtw_k5mean_ts}
                \end{subfigure}
                ~
                % Subfigure 2
                \begin{subfigure}[b]{0.51\textwidth}
                    \centering
                    \includegraphics[width=\textwidth, trim={0.2cm 0.4cm 0.2cm 1.4cm}]{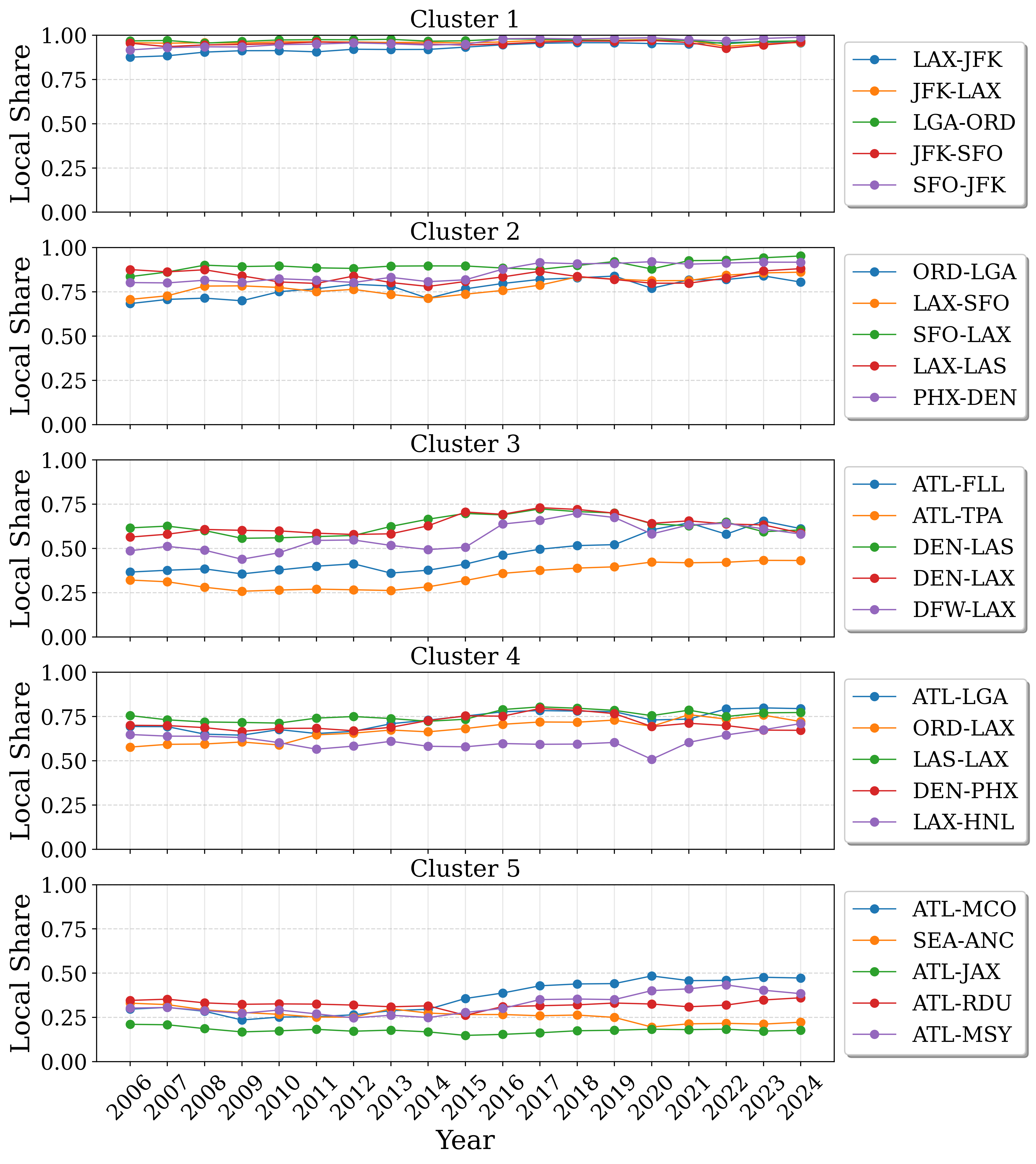}
                    \caption{Top total pax O\&D pairs for each cluster.}
                    \label{fig:dtw_k5_representative_plot}
                \end{subfigure}
                % \vspace{0.5em} 
                \caption{Clustering visualization for hierarchical clustering with DynTW ($k=5$).}
                \label{fig:dtw_k5}
            \end{figure}

            % part 1 - original magnitude with three subplots
            \paragraph{Original Magnitude Data Clustering Overview} The clustering results were visualized from multiple perspectives to provide an intuitive understanding of the local share dynamics across the O\&D pairs. Figure~\ref{fig:dtw_k5} presents the clustering outcomes from Hierarchical Clustering with DynTW using the original magnitude of the local share time series. In Figure~\ref{fig:dtw_k5mean_ts}, the average local share values for each cluster are displayed, accompanied by 95\% confidence intervals to illustrate the variability within clusters. This visualization highlights the distinct magnitude differences across clusters, revealing hierarchical levels of local share. 
    
            The results indicate that this clustering method primarily groups O\&D pairs based on the magnitude of the local share, with shape-based similarity as a secondary role. As depicted in Figure \ref{fig:dtw_k5_representative_plot}, the top five O\&D pairs with the highest total passengers in each cluster exhibit different local share patterns. For instance, Cluster 1 consists of O\&D pairs with relatively stable local shares over time, predominantly involving large airports on the East or West Coast. These airports, located in major metropolitan areas with large populations, enjoy a substantial number of local passengers. In contrast, other clusters with lower average local share values include large airports situated in the central regions of the country, e.g., ATL, DEN, and ORD. These airports serve as major transfer hubs connecting the East and West Coasts and consequently handling a large volume of transfer passengers in the U.S. domestic market. It is important to note that the clustering numbers are merely index labels without intrinsic meaning, as they result from unsupervised clustering.

            % part 2 - standardized magnitude with four subplots

\begin{figure}[ht!]
                \centering
                % Subfigure 1
                \begin{subfigure}[b]{0.42\textwidth}
                    \centering
                    \includegraphics[width=\textwidth, trim={0cm 0.5cm 0cm 2cm}, clip]{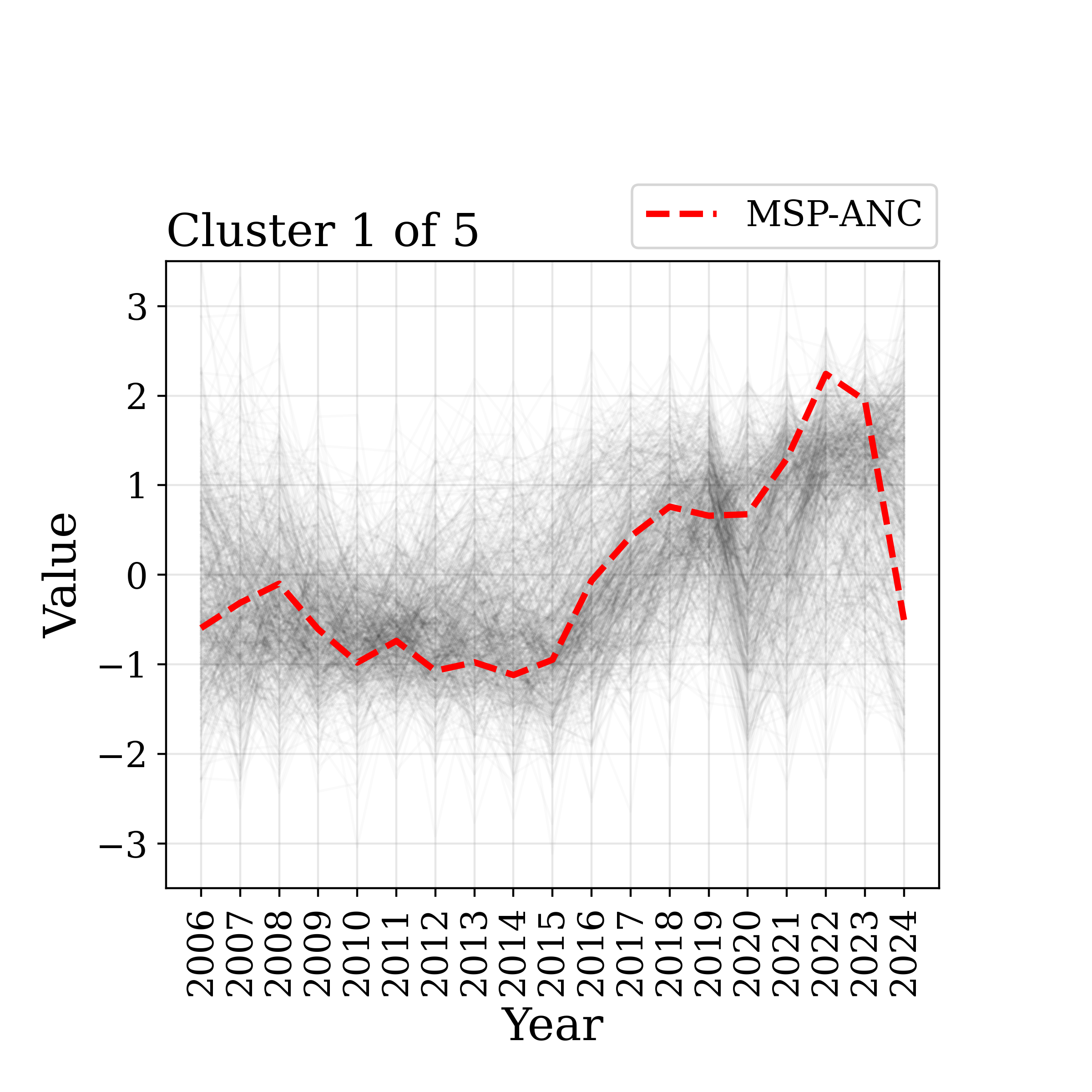}
                    \caption{$k$-Shape.}
                    \label{fig:kshape_k5_mspanc}
                \end{subfigure}
                ~
                % Subfigure 2
                \begin{subfigure}[b]{0.42\textwidth}
                    \centering
                    \includegraphics[width=\textwidth, trim={0cm 0.5cm 0cm 2cm}, clip]{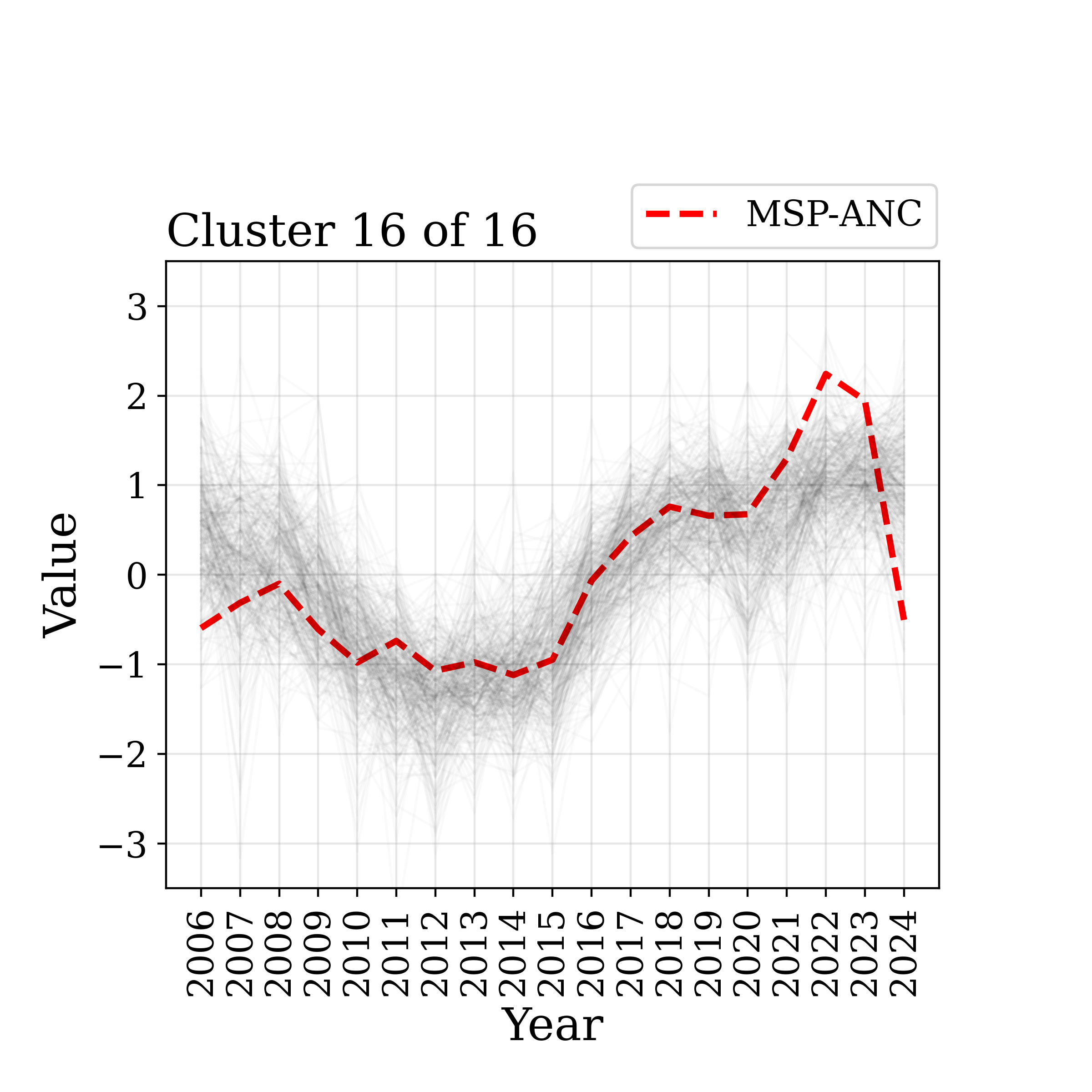}
                    \caption{SOM.}
                    \label{fig:som_k16_mspanc}
                \end{subfigure}
                \\
                \vspace{0.5em}
                % Subfigure 3
                \begin{subfigure}[b]{0.42\textwidth}
                    \centering
                    \includegraphics[width=\textwidth, trim={0cm 0.5cm 0cm 2cm}, clip]{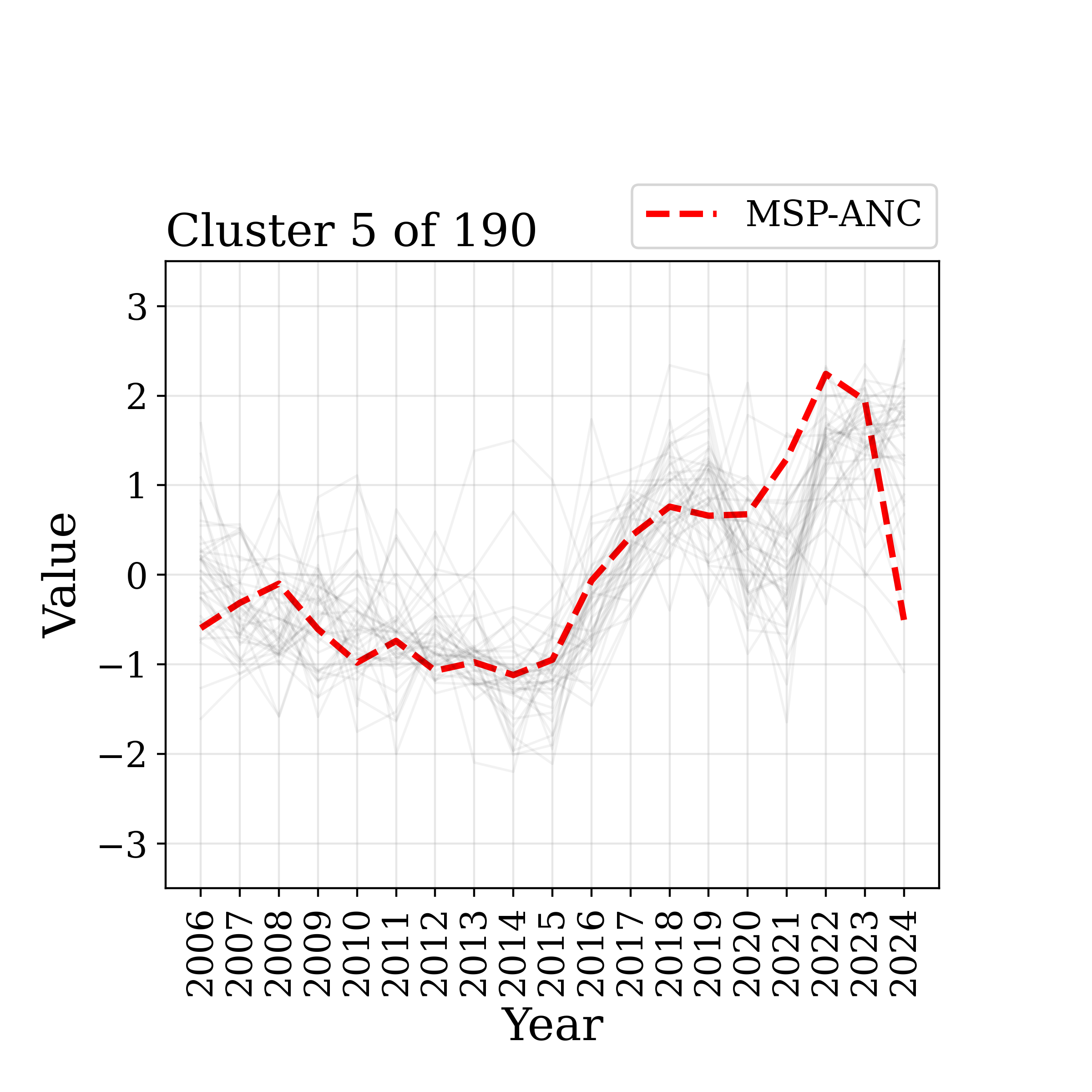}
                    \caption{SBD with AP.}
                    \label{fig:sbd_k190_mspanc}
                \end{subfigure}
                ~
                % Subfigure 4
                \begin{subfigure}[b]{0.42\textwidth}
                    \centering
                    \includegraphics[width=\textwidth, trim={0cm 0.5cm 0cm 2cm}, clip]{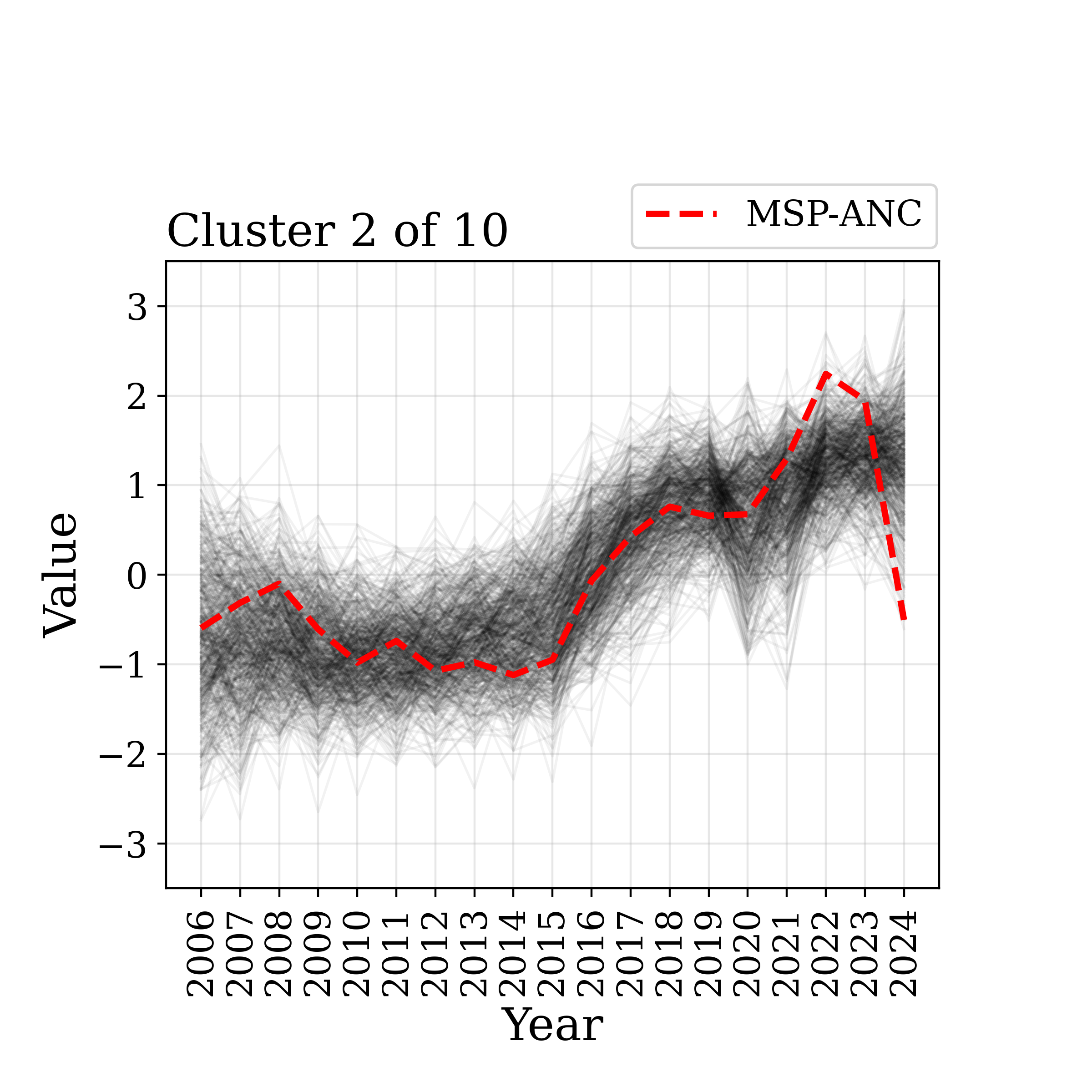}
                    \caption{DBA with GMM.}
                    \label{fig:gmm_k10_mspanc}
                \end{subfigure}
                % \vspace{0.5em}
                \caption{Clustering visualization - clusters including MSP $\rightarrow$ ANC with standardized local share.}
                \label{fig:msp_anc}
            \end{figure}

\begin{figure}[ht!]
                \centering
                % Subfigure 1
                \begin{subfigure}[b]{0.42\textwidth}
                    \centering
                    \includegraphics[width=\textwidth, trim={0cm 0.5cm 0cm 2cm}, clip]{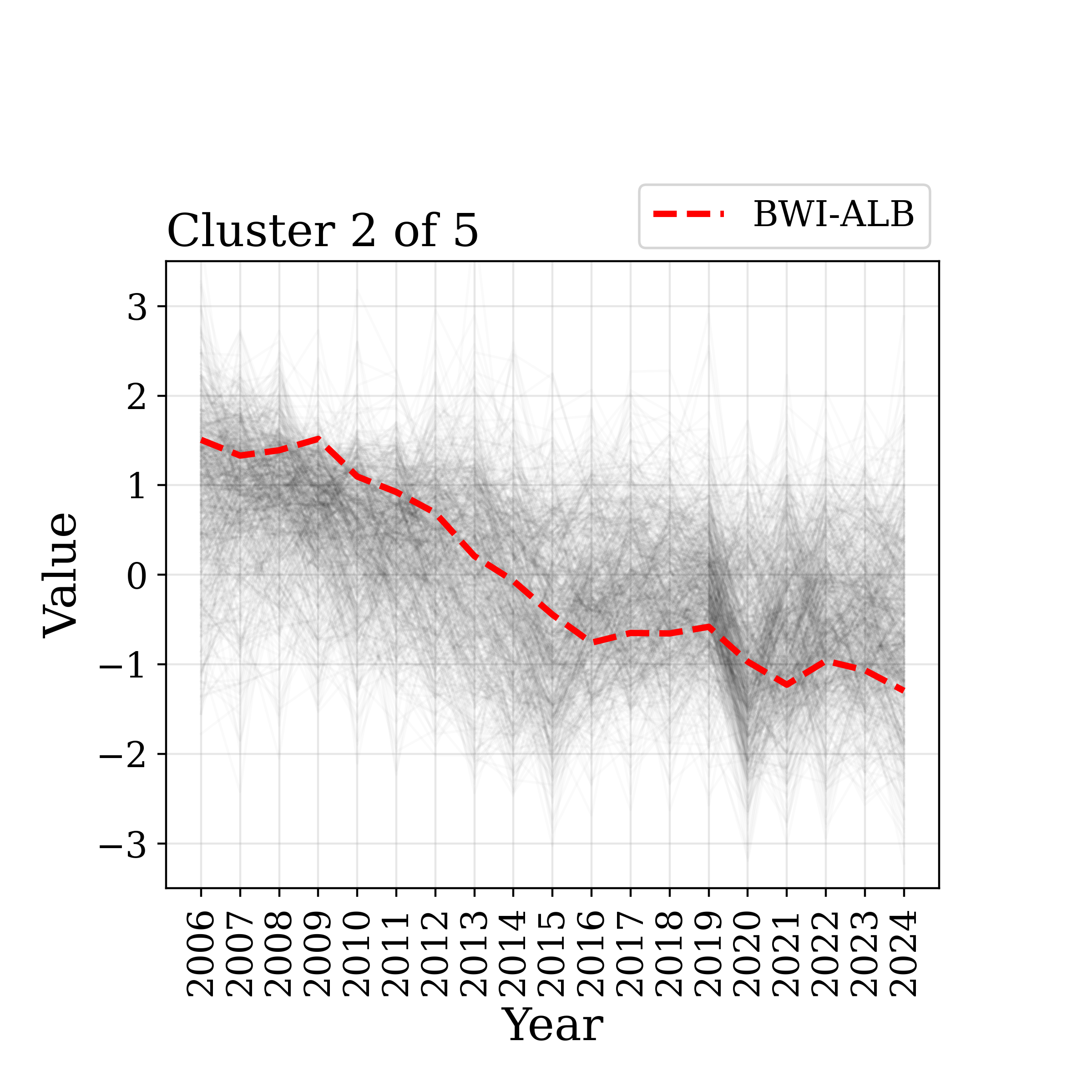}
                    \caption{$k$-Shape.}
                    \label{fig:kshape_k5_bwialb}
                \end{subfigure}
                ~
                % Subfigure 2
                \begin{subfigure}[b]{0.42\textwidth}
                    \centering
                    \includegraphics[width=\textwidth, trim={0cm 0.5cm 0cm 2cm}, clip]{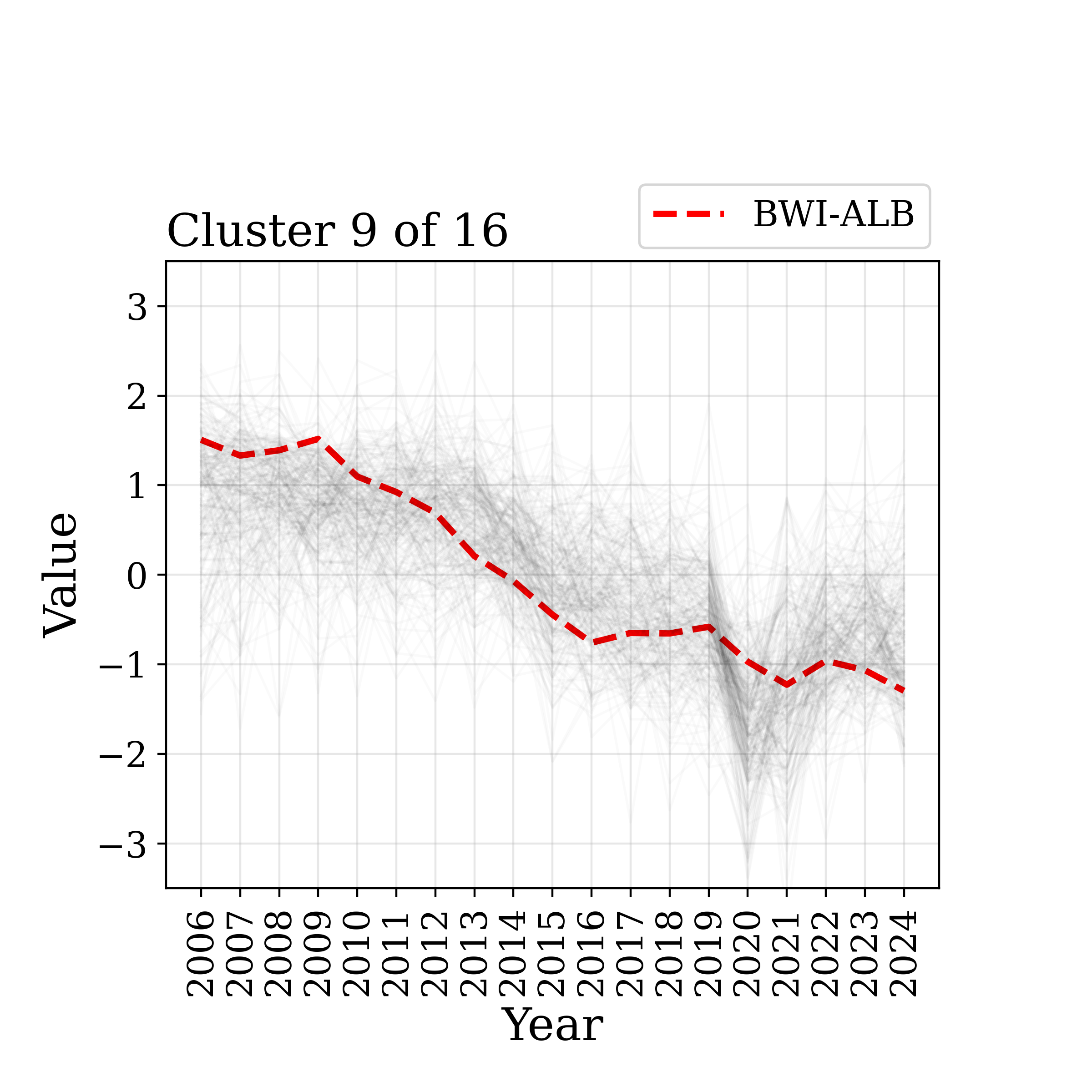}
                    \caption{SOM.}
                    \label{fig:som_k16_bwialb}
                \end{subfigure}
                \\
                \vspace{0.5em}
                % Subfigure 3
                \begin{subfigure}[b]{0.42\textwidth}
                    \centering
                    \includegraphics[width=\textwidth, trim={0cm 0.5cm 0cm 2cm}, clip]{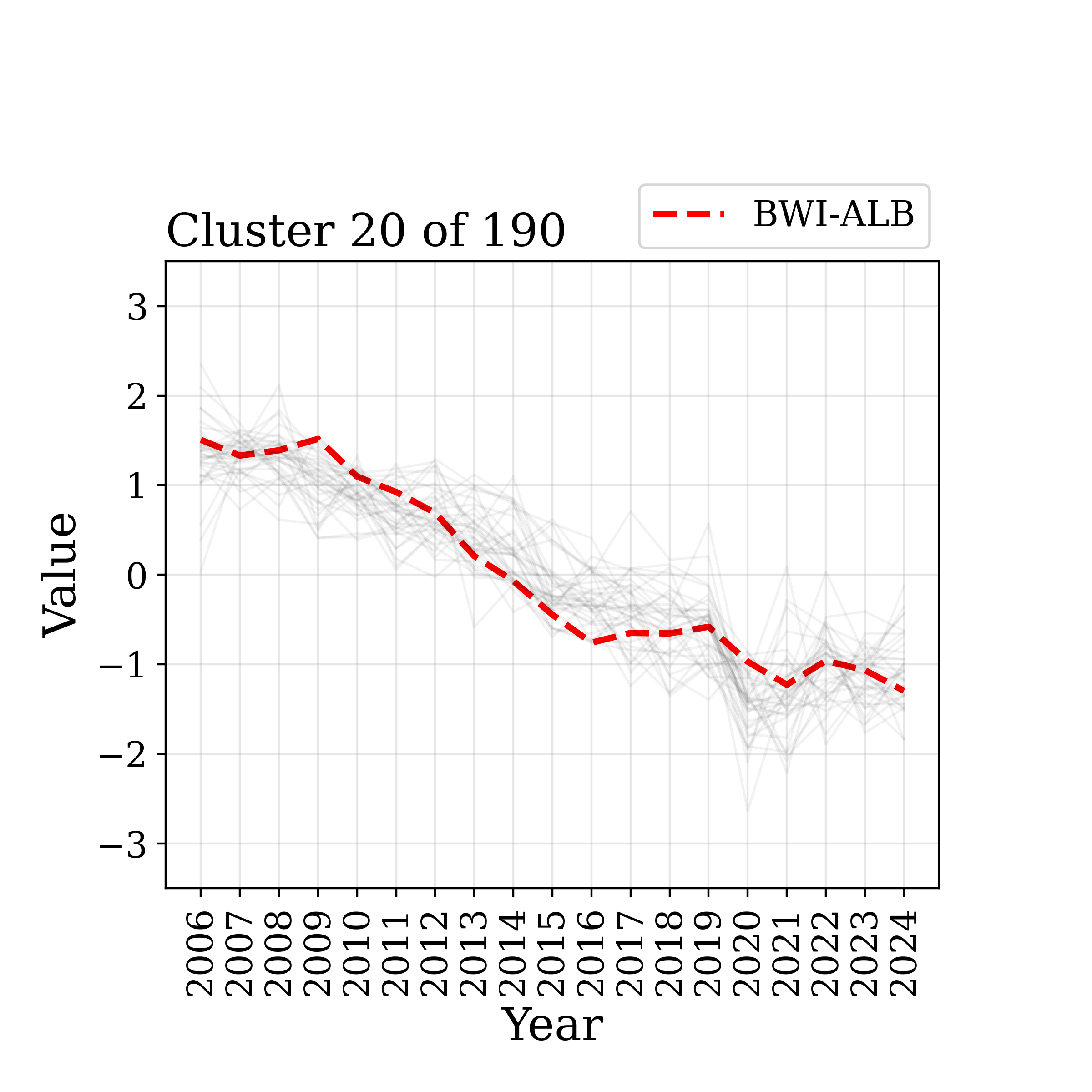}
                    \caption{SBD with AP.}
                    \label{fig:sbd_k190_bwialb}
                \end{subfigure}
                ~
                % Subfigure 4
                \begin{subfigure}[b]{0.42\textwidth}
                    \centering
                    \includegraphics[width=\textwidth, trim={0cm 0.5cm 0cm 2cm}, clip]{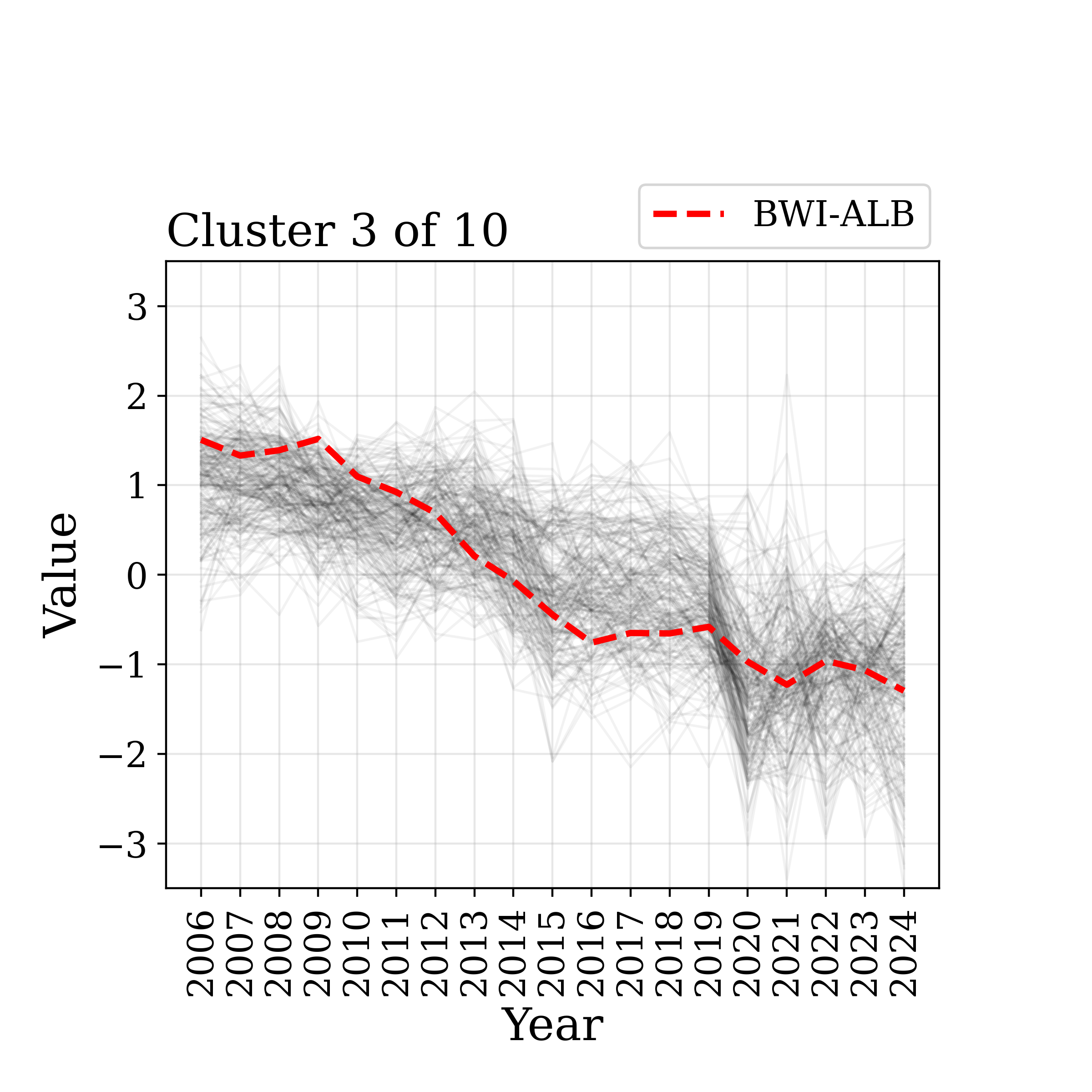}
                    \caption{DBA with GMM.}
                    \label{fig:gmm_k10_bwialb}
                \end{subfigure}
                % \vspace{0.5em}
                \caption{Clustering visualization - cluster including BWI $\rightarrow$ ALB with standardized local share.}
                \label{fig:bwi_alb}
            \end{figure}

\begin{figure}[ht!]
                \centering
                % Subfigure 1
                \begin{subfigure}[b]{0.42\textwidth}
                    \centering
                    \includegraphics[width=\textwidth, trim={0cm 0.5cm 0cm 2cm}, clip]{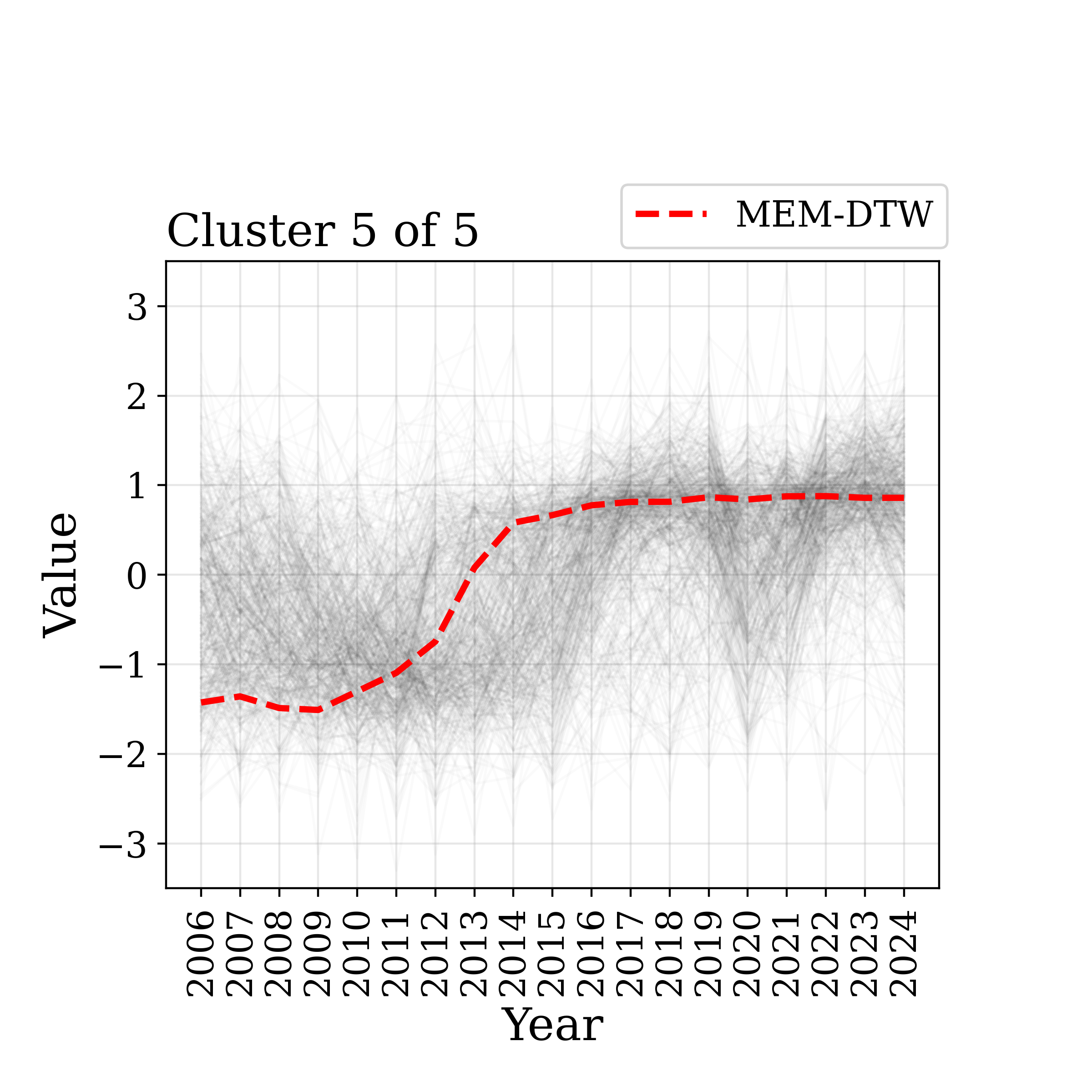}
                    \caption{$k$-Shape.}
                    \label{fig:kshape_k5_memdtw}
                \end{subfigure}
                ~
                % Subfigure 2
                \begin{subfigure}[b]{0.42\textwidth}
                    \centering
                    \includegraphics[width=\textwidth, trim={0cm 0.5cm 0cm 2cm}, clip]{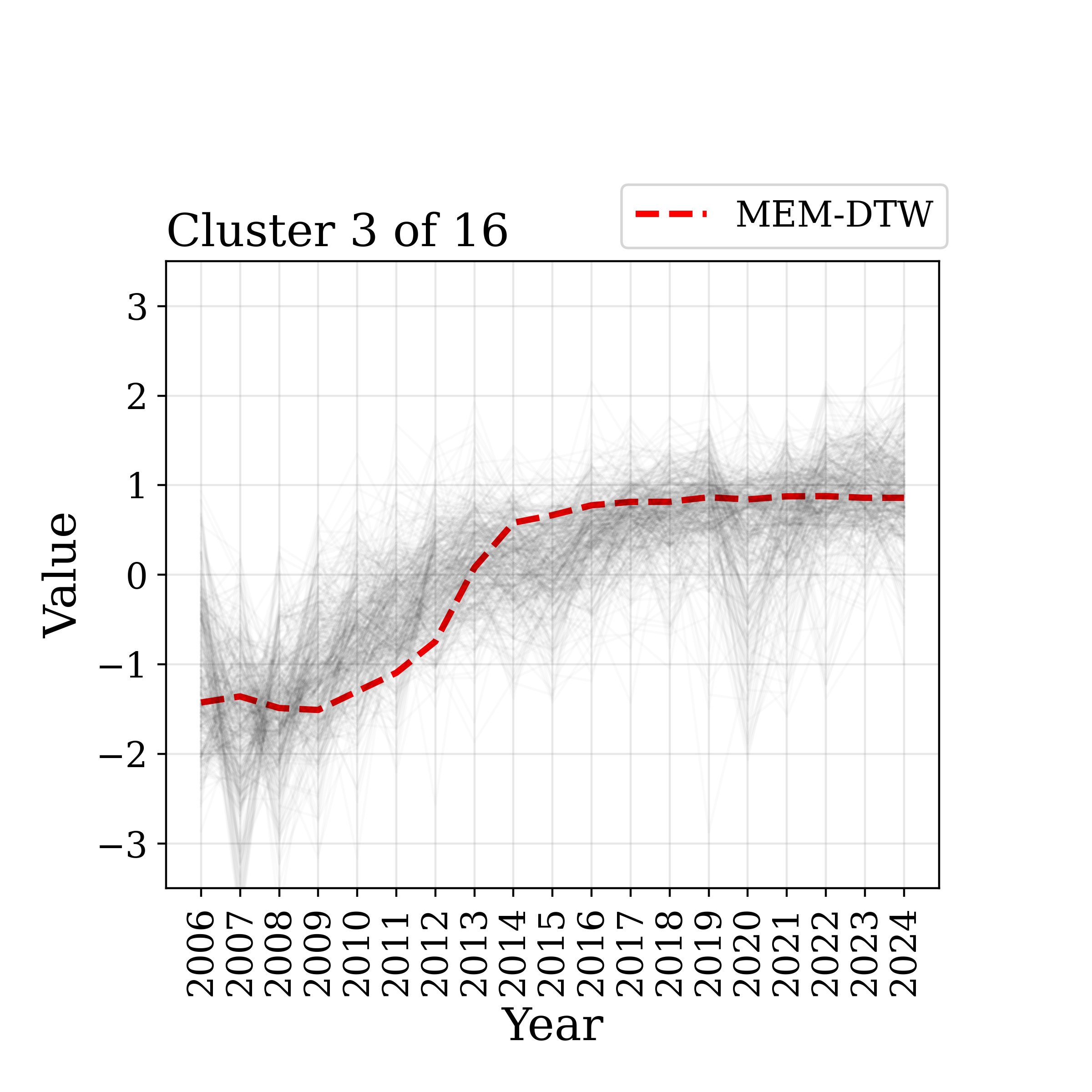}
                    \caption{SOM.}
                    \label{fig:som_k16_memdtw}
                \end{subfigure}
                \\
                \vspace{0.5em}
                % Subfigure 3
                \begin{subfigure}[b]{0.42\textwidth}
                    \centering
                    \includegraphics[width=\textwidth, trim={0cm 0.5cm 0cm 2cm}, clip]{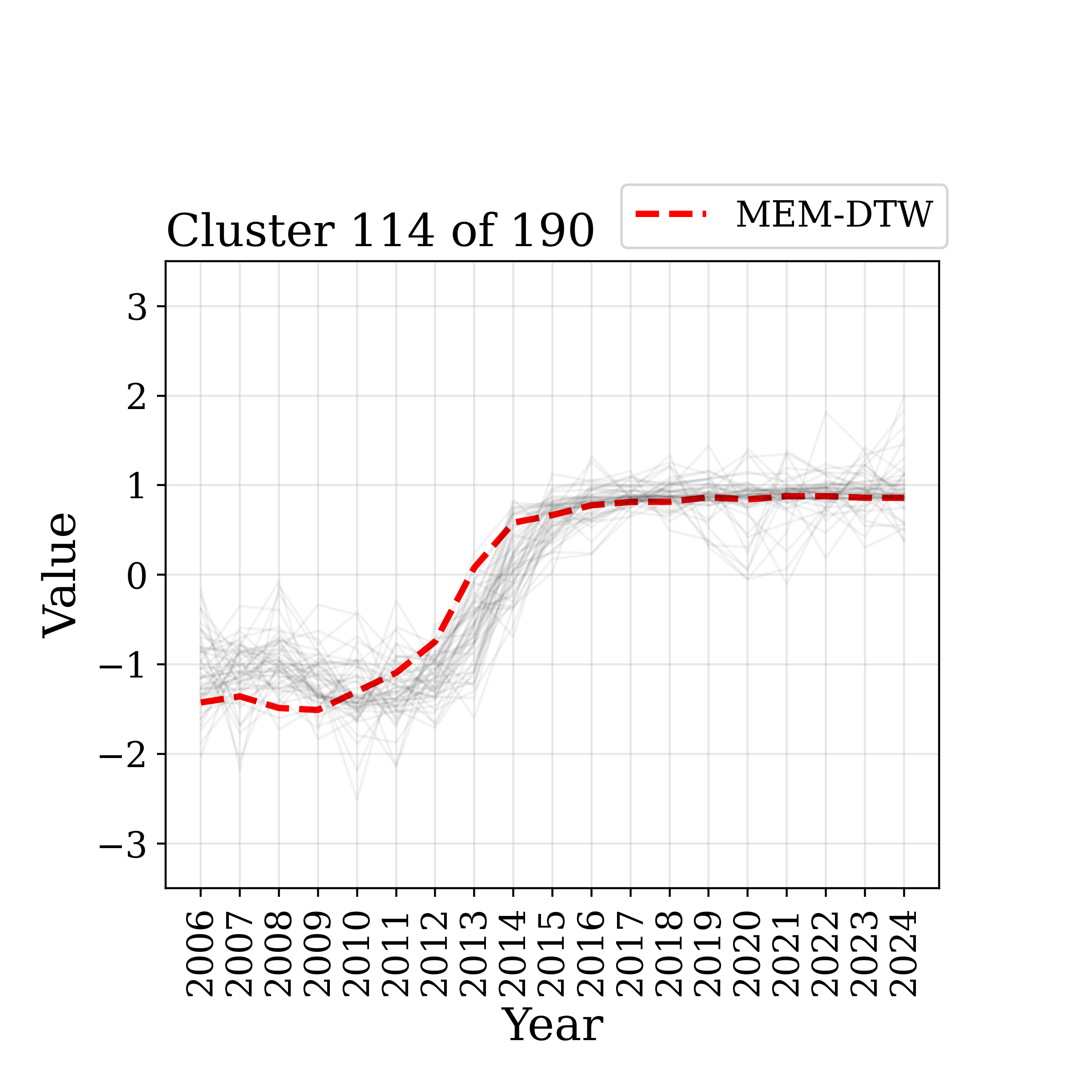}
                    \caption{SBD with AP.}
                    \label{fig:sbd_k190_memdtw}
                \end{subfigure}
                ~
                % Subfigure 4
                \begin{subfigure}[b]{0.42\textwidth}
                    \centering
                    \includegraphics[width=\textwidth, trim={0cm 0.5cm 0cm 2cm}, clip]{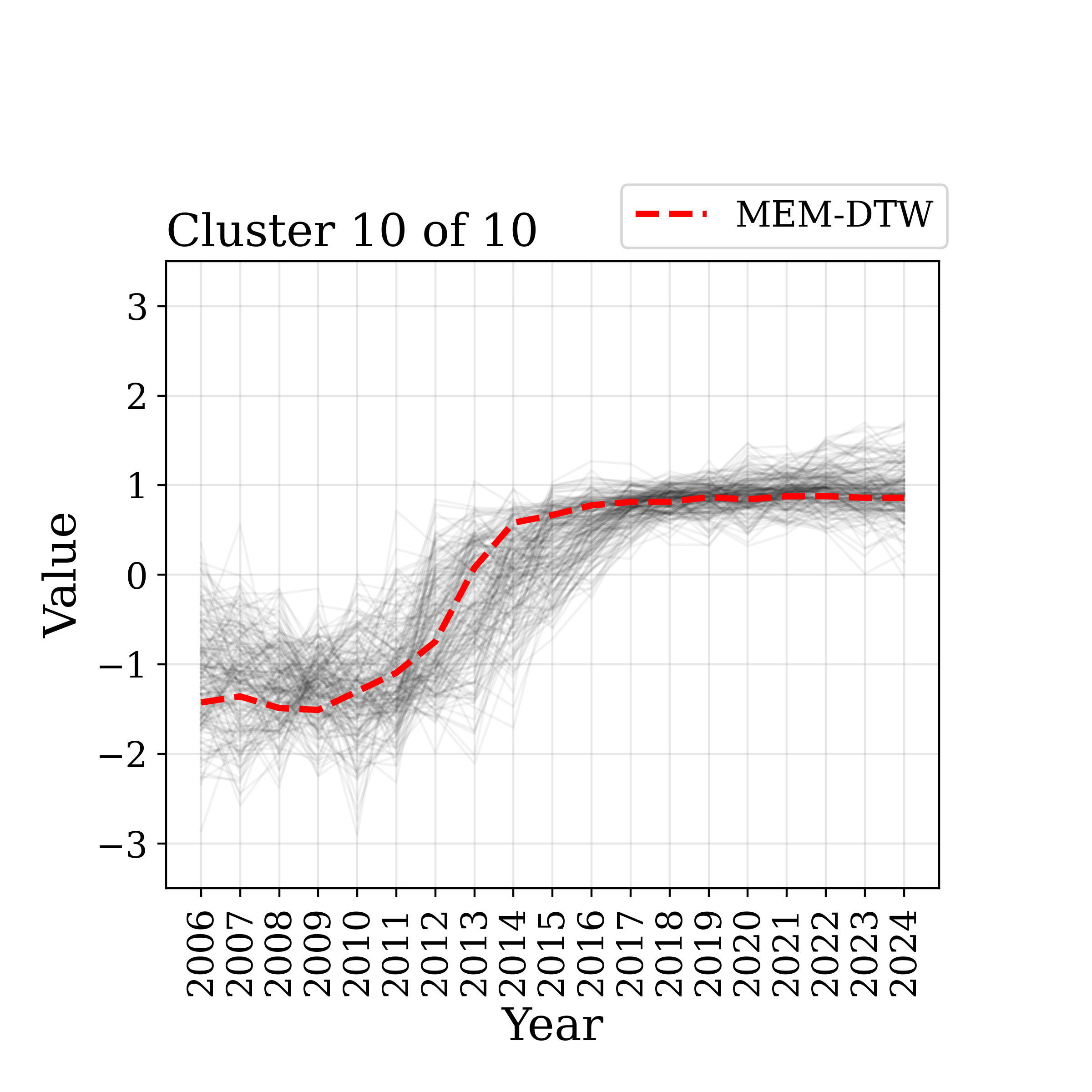}
                    \caption{DBA with GMM.}
                    \label{fig:gmm_k10_memdtw}
                \end{subfigure}
                % \vspace{0.5em}
                \caption{Clustering visualization - cluster including MEM $\rightarrow$ DTW with standardized local share.}
                \label{fig:mem_dtw}
            \end{figure}

            \paragraph{Standardized Data Clustering Overview} To further illustrate the clustering outcomes focusing on the shape and trend of the local share time series, a series of figures derived from the standardized local share time series are presented. These figures showcase the clustering results from $k$-Shape, SOM, SBD with AP, and DBA with GMM, each displayed in a separate subplot. Three representative O\&D pairs are highlighted in the plots: MSP-ANC in Section \ref{sec:Analysis_MSPANC}, BWI-ALB in Section \ref{sec:Analysis_BWIALB}, and MEM-DTW in Section \ref{sec:Analysis_MEMDTW}. Figure \ref{fig:msp_anc} presents the O\&D clustering results including MSP-ANC and Figure \ref{fig:bwi_alb} and Figure \ref{fig:mem_dtw} display the O\&D clustering results including BWI-ALB and MEM-DTW, respectively. In these clustering figures, the gray lines represent the local share time series of all O\&D pairs within the respective cluster, while the red lines depict these selected O\&D pairs described above. The cluster number in each subplot is labeled with a nominal index, and the total number of clusters for each method is also indicated.

\begin{figure}[ht!]
                \centering
                % Subfigure 1
                \begin{subfigure}[b]{0.42\textwidth}
                    \centering
                    \includegraphics[width=\textwidth, trim={0cm 0.5cm 0cm 2cm}, clip]{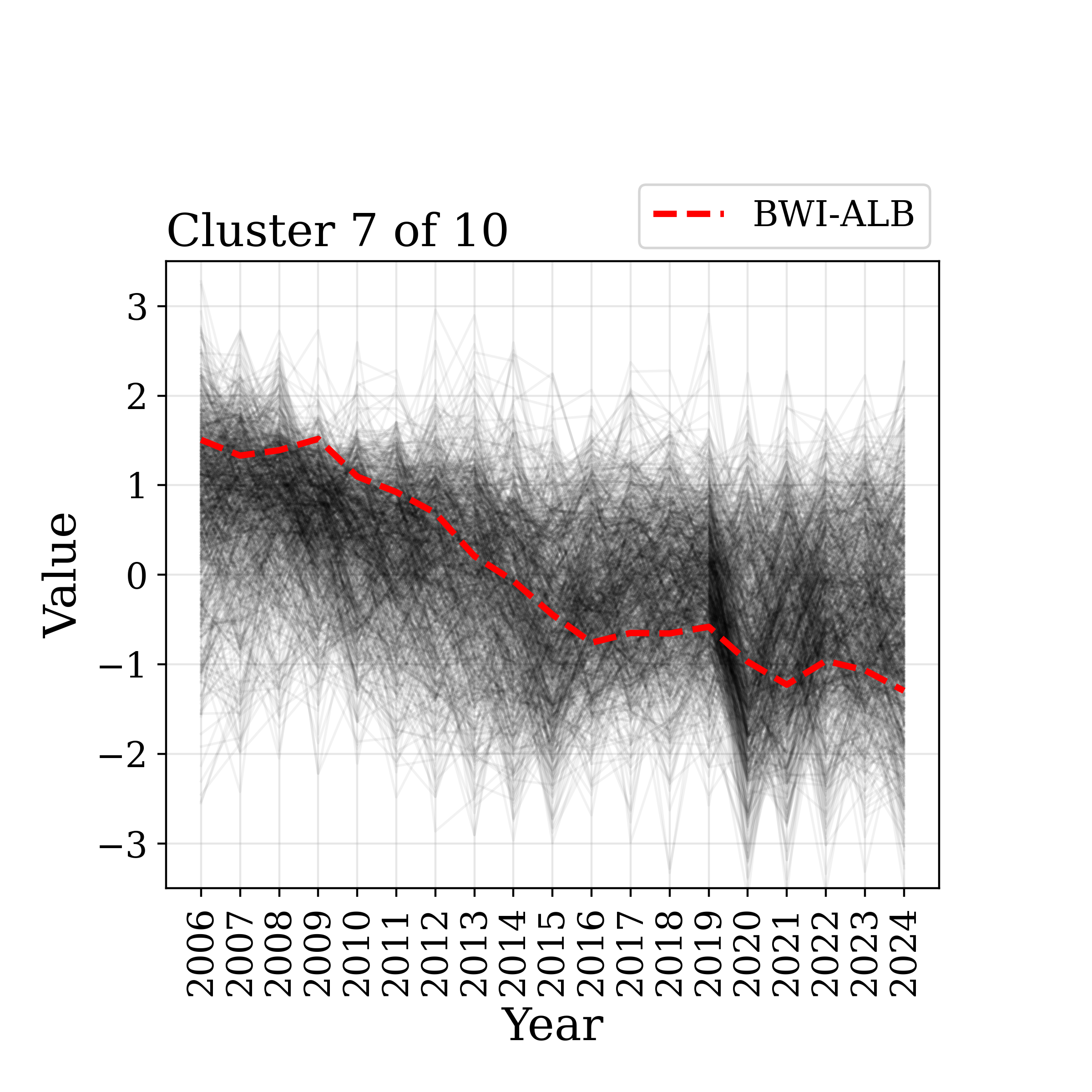}
                    \caption{O\&D cluster including BWI-ALB}
                    \label{fig:sbd_k190_bwialb_hc}
                \end{subfigure}
                ~
                % Subfigure 2
                \begin{subfigure}[b]{0.42\textwidth}
                    \centering
                    \includegraphics[width=\textwidth, trim={0cm 0.5cm 0cm 2cm}, clip]{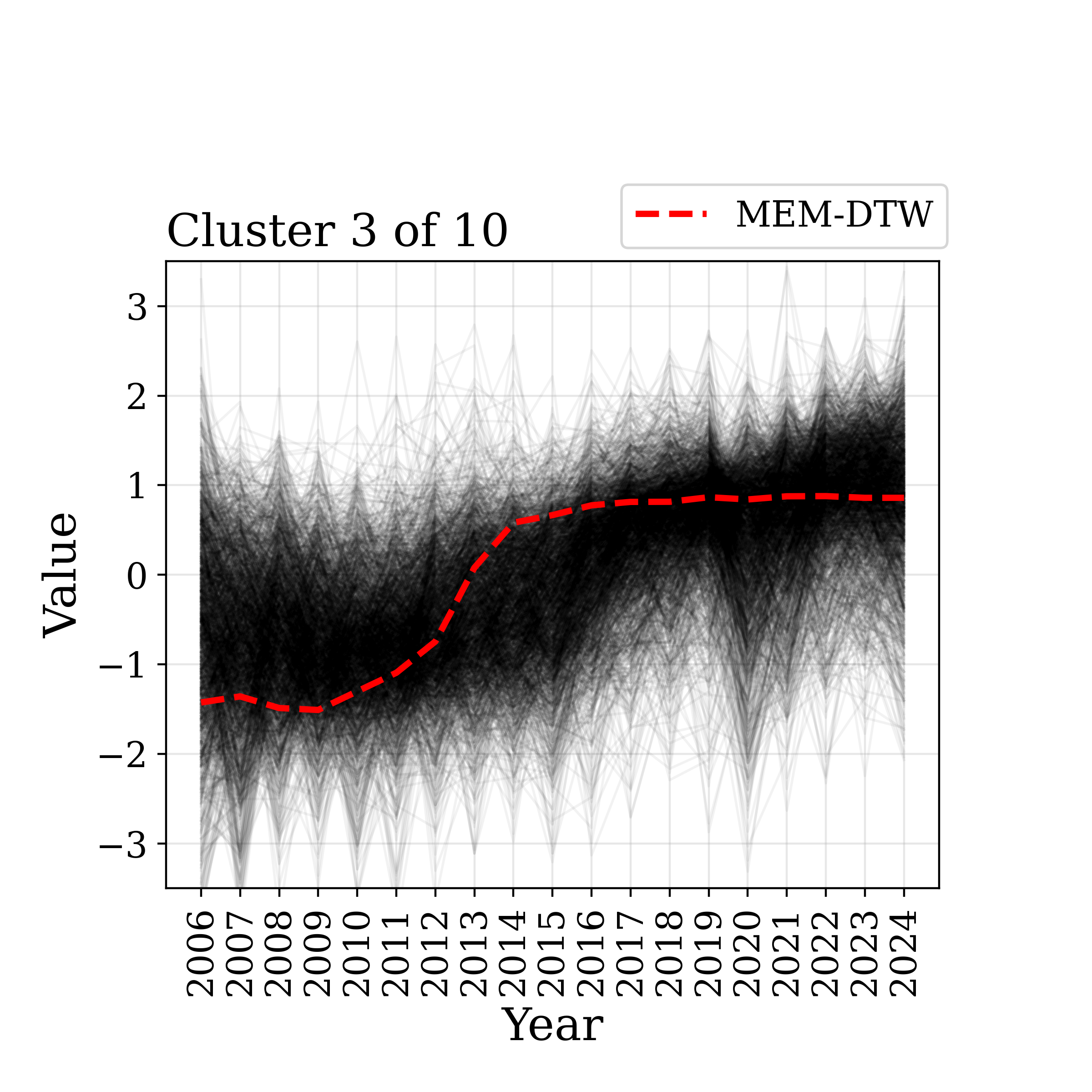}
                    \caption{O\&D cluster including MEM $\rightarrow$ DTW}
                    \label{fig:sbd_k190_memdtw_hc}
                \end{subfigure}
                % \vspace{0.5em}
                \caption{Clustering visualization from one two-Step approach for selected O\&D pairs at $k=10$.}
                \label{fig:two_step_clustering}
            \end{figure}

            Generally, a larger number of clusters results in more precise clustering, yielding a more detailed representation. For instance, the SBD with AP method using $k=190$ provides more granular distinctions than the $k$-Shape method with $k=5$. In addition, the SOM approach (Figure~\ref{fig:som_k16_mspanc}) demonstrated its capacity to capture essential time-series characteristics with a relatively small number of clusters (i.e., $k=16$), making it suitable for preliminary system-wide analyses. Two-step clustering approaches, such as combining SBD with AP and then applying HC, were also explored to reduce the complexity of the clustering results. Figure \ref{fig:two_step_clustering} illustrates these two-step clustering outcomes for the BWI-ALB and MEM-DTW O\&D pairs under a reduced $k=10$ setting. Compared to larger $k$ values, these aggregated clusters highlighted broad directional movements, while the more granular clustering identified subtle step-wise decreases (e.g., BWI-ALB) or plateaus (e.g., MEM-DTW). Despite differences in clustering detail, all methods effectively grouped O\&D pairs with similar local share dynamics, indicating step-wise increases, decreases, or stable trends, as displayed in Figures~\ref{fig:msp_anc},~\ref{fig:bwi_alb}, and~\ref{fig:mem_dtw}.

\begin{figure}[ht!]
                \centering
                % Subfigure 1
                \begin{subfigure}[b]{0.42\textwidth}
                    \centering
                    \includegraphics[width=\textwidth, trim={0cm 0.5cm 0cm 1cm},clip]{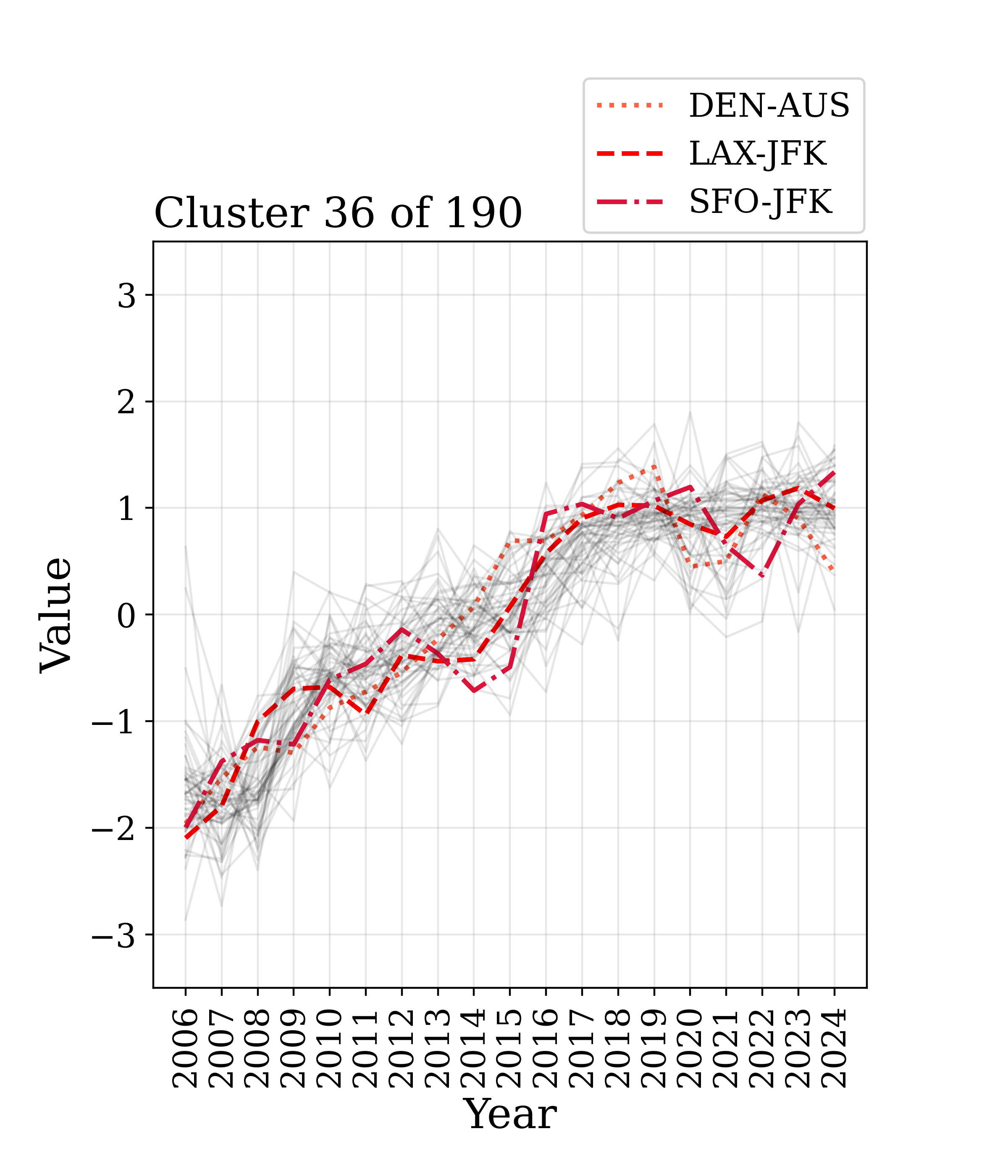}
                    \caption{Cluster based on standardized local share values.}
                    \label{fig:sbd_k190_denaus_original}
                \end{subfigure}
                ~
                % Subfigure 2
                \begin{subfigure}[b]{0.42\textwidth}
                    \centering
                    \includegraphics[width=\textwidth, trim={0cm 0.5cm 0cm 1cm},clip]{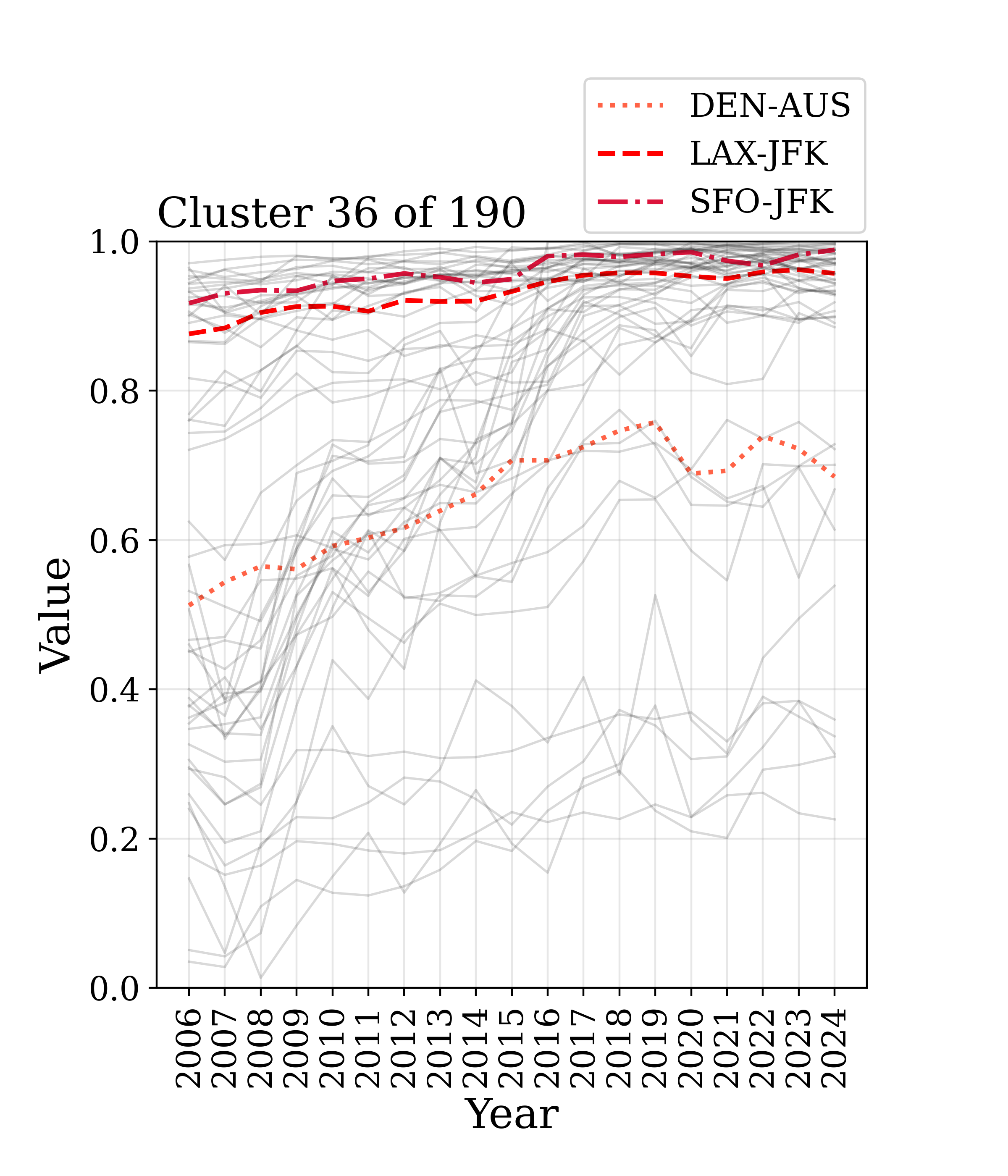}
                    \caption{Same cluster displayed in original magnitude.}
                    \label{fig:sbd_k190_denaus__original}
                \end{subfigure}
                % \vspace{0.5em}
                \caption{Clustering visualization - standardized clustering results in the original magnitude range.}
                \label{fig:original_magnitude_DENAUS}
            \end{figure}

            The standardized local share values enabled an additional interpretive layer by magnifying patterns less discernible in the original magnitude. For example, Cluster 1 in Figure~\ref{fig:dtw_k5_representative_plot} contained O\&D pairs like SFO (San Francisco International Airport, San Francisco, CA) $\rightarrow$ JFK (John F. Kennedy International Airport, Queens, NY) and LAX (Los Angeles International Airport, Los Angeles, CA) $\rightarrow$ JFK, both maintaining local share near 1. By contrast, in the standardized clustering results (Cluster 36 via SBD with AP method), O\&D pairs such as DEN $\rightarrow$ AUS (Austin-Bergstrom International Airport, Austin, TX) that ranges from 0.5 to 0.7 in original local share magnitude, were grouped with these transcontinental routes, as exhibited in Figure~\ref{fig:original_magnitude_DENAUS}. This observation suggests that local share trends may synchronize across O\&D pairs with differing magnitudes, indicating shared factors driving these progressive fluctuations. In the case of Cluster 36 (SBD with AP method), all of the O\&D pairs exhibited similar steadily increasing trends, encompassing coast-to-coast routes (e.g., SFO $\rightarrow$ JFK, LAX $\rightarrow$ JFK), and north-south connections, such as MKE (Milwaukee Mitchell International Airport, Milwaukee, WI) $\rightarrow$ TPA. These shared geographic and metropolitan attributes likely contributed to similar demand growth, reflected in their convergent trends.

\begin{figure}[ht!]
                \centering
                % Subfigure 1
                \begin{subfigure}[b]{0.42\textwidth}
                    \centering
                    \includegraphics[width=\textwidth, trim={0cm 0.5cm 0cm 2cm},clip]{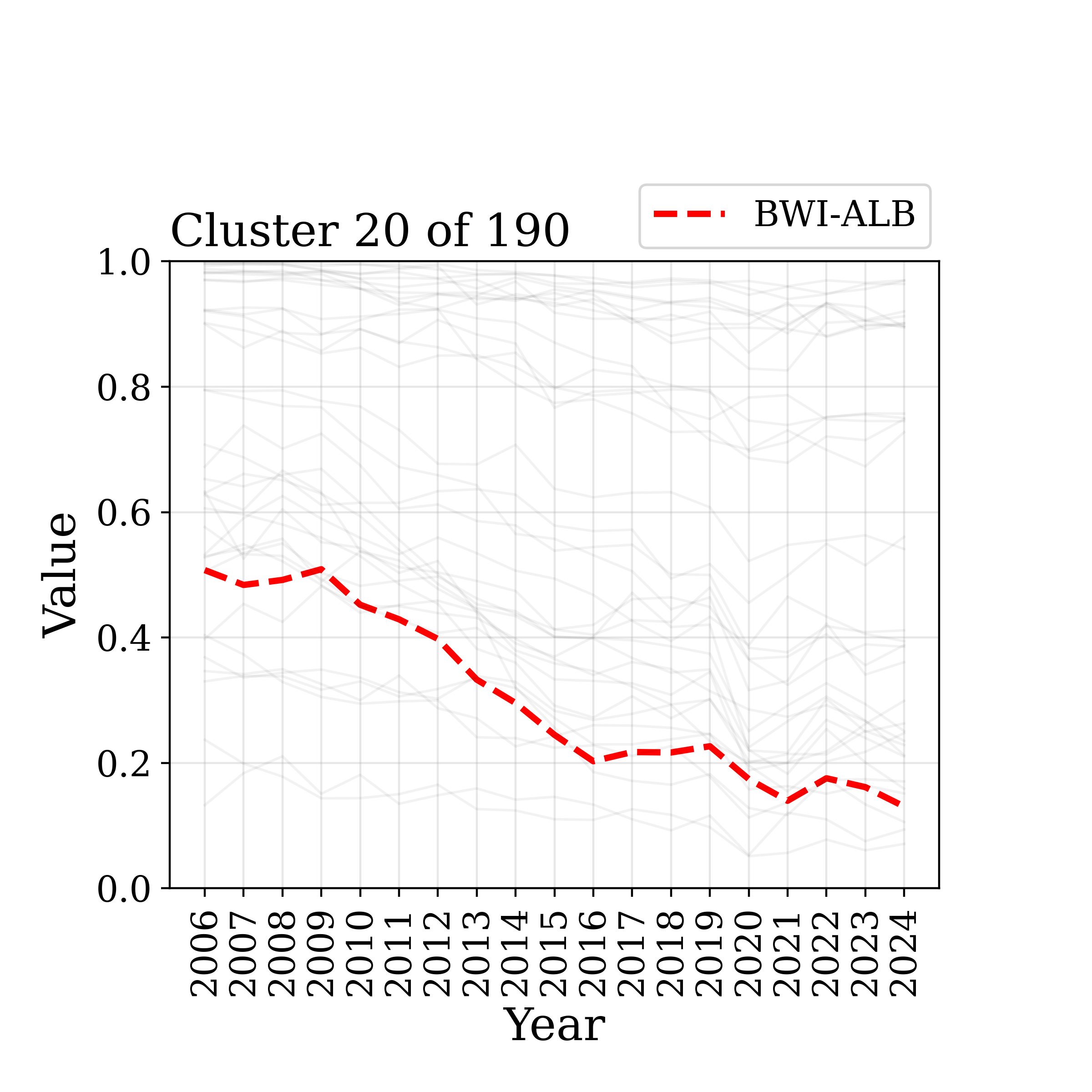}
                    \caption{O\&D cluster including BWI-ALB.}
                    \label{fig:sbd_k190_bwialb_original}
                \end{subfigure}
                ~
                % Subfigure 2
                \begin{subfigure}[b]{0.42\textwidth}
                    \centering
                    \includegraphics[width=\textwidth, trim={0cm 0.5cm 0cm 2cm},clip]{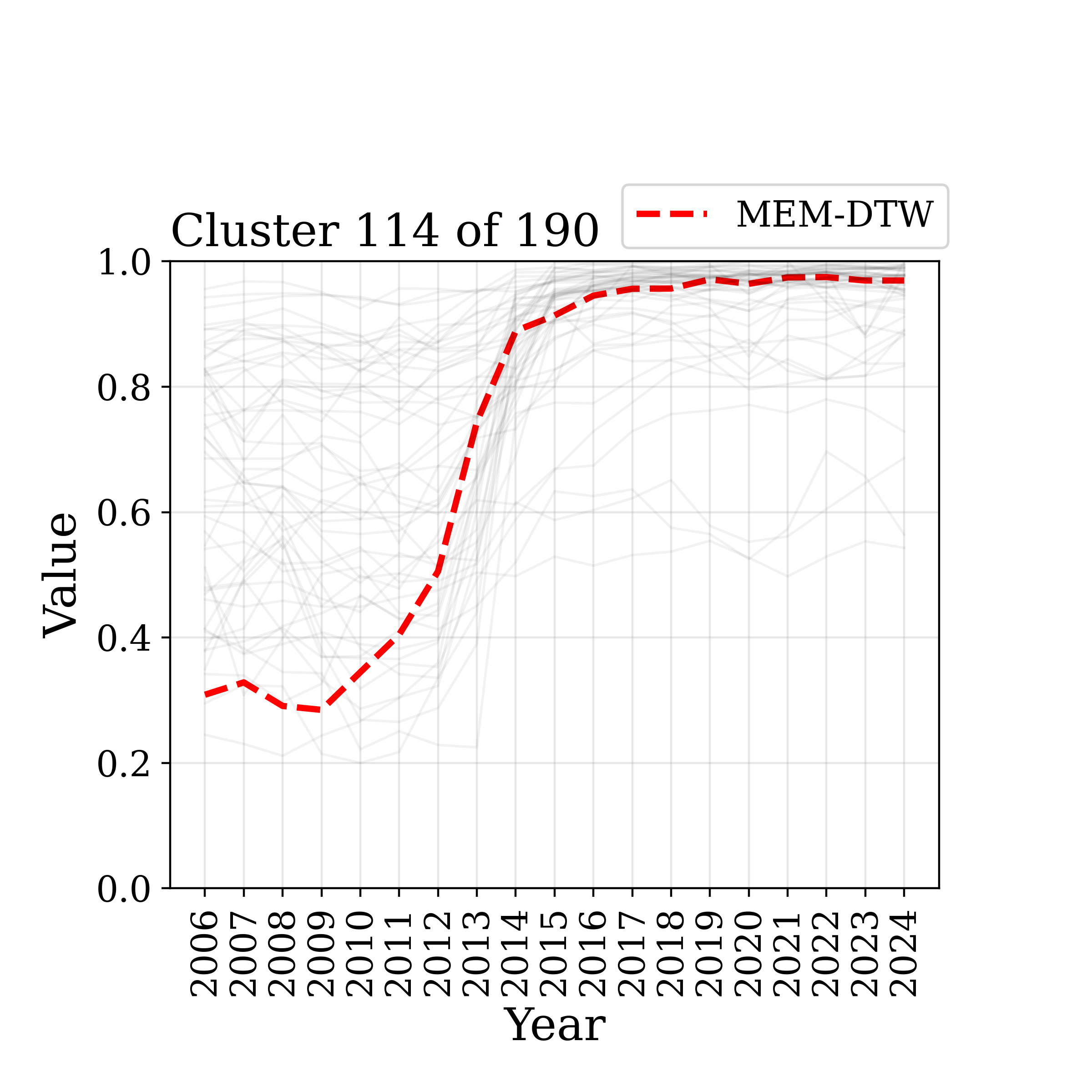}
                    \caption{O\&D cluster including MEM-DTW.}
                    \label{fig:sbd_k190_memdtw_original}
                \end{subfigure}
                % \vspace{0.5em}
                \caption{Clustering visualization - clusters including BWI $\rightarrow$ ALB and MEM $\rightarrow$ DTW with original local share.}
                \label{fig:original_magnitude}
            \end{figure}

            Figures \ref{fig:sbd_k190_bwialb_original} and \ref{fig:sbd_k190_memdtw_original} further present the same clusters (via SBD with AP method) in original magnitude view based on standardized value clustering results, which were shown previously in Figures \ref{fig:sbd_k190_bwialb} and \ref{fig:sbd_k190_memdtw}, respectively. This comparison reveals that O\&D pairs grouped by standardized patterns may occupy varied absolute local share levels. Focusing on the cluster including BWI $\rightarrow$ ALB, this cluster, generated by standardized values, contains O\&D pairs spanning the entire local share range. Pairs with higher initial local share appeared more resilient to declines, while those starting lower were more vulnerable to overarching market shifts. Additionally, a considerable number of O\&D pairs in this cluster connect BWI to other airports, such as BDL (Bradley International Airport, Windsor Locks, CT), BUF (Buffalo Niagara International Airport, Buffalo, NY), and CLE (Cleveland Hopkins International Airport, Cleveland, OH). As noted in Section \ref{sec:Analysis}, adjustments in Southwest Airlines' operational strategy at BWI influenced local share trends across multiple O\&D pairs, causing them to behave similarly despite differing baseline local share levels. This observation underscores the clustering method's capability to identify common factors driving shared temporal patterns, regardless of initial local share magnitudes.

            A comparable pattern was observed for MEM $\rightarrow$ DTW as well, which displayed an increasing trend before reaching a plateau. Standardized results indicated that these O\&D pairs collectively transitioned from negative to positive standardized values (i.e., $Z$-scores), while some experienced a more pronounced rise when viewed in the original magnitude. In this cluster, they were mostly influenced by network adjustments, such as Delta's realignment of MEM from a transfer hub to a regional airport, ultimately leading to increased local share stability after 2015. By examining these O\&D pair clustering by standardized values, and displaying them in their original magnitude, the analysis isolated common trend dynamics, illustrating how airline strategic decisions can produce synchronized local share evolution across diverse markets.

            \paragraph{Distribution of Cluster Membership} Examining the distribution of O\&D pairs across clusters provides additional insights into the nature of the clustering outcomes. A more balanced distribution indicates that multiple clusters capture recurring and broad shared patterns, while a skewed distribution may reflect either the presence of rare but meaningful patterns or an over-segmentation of the data. To quantify the distribution's balance, the Gini coefficient was introduced, ranging from 0 (perfect equality) to 1 (perfect inequality)~\cite{bendel1989comparison}. 

            \begin{figure}[ht!]
                \centering
                % Subfigure 1
                \begin{subfigure}[b]{0.45\textwidth}
                    \centering
                    \includegraphics[width=\textwidth, trim={0cm 0cm 0cm 0cm}]{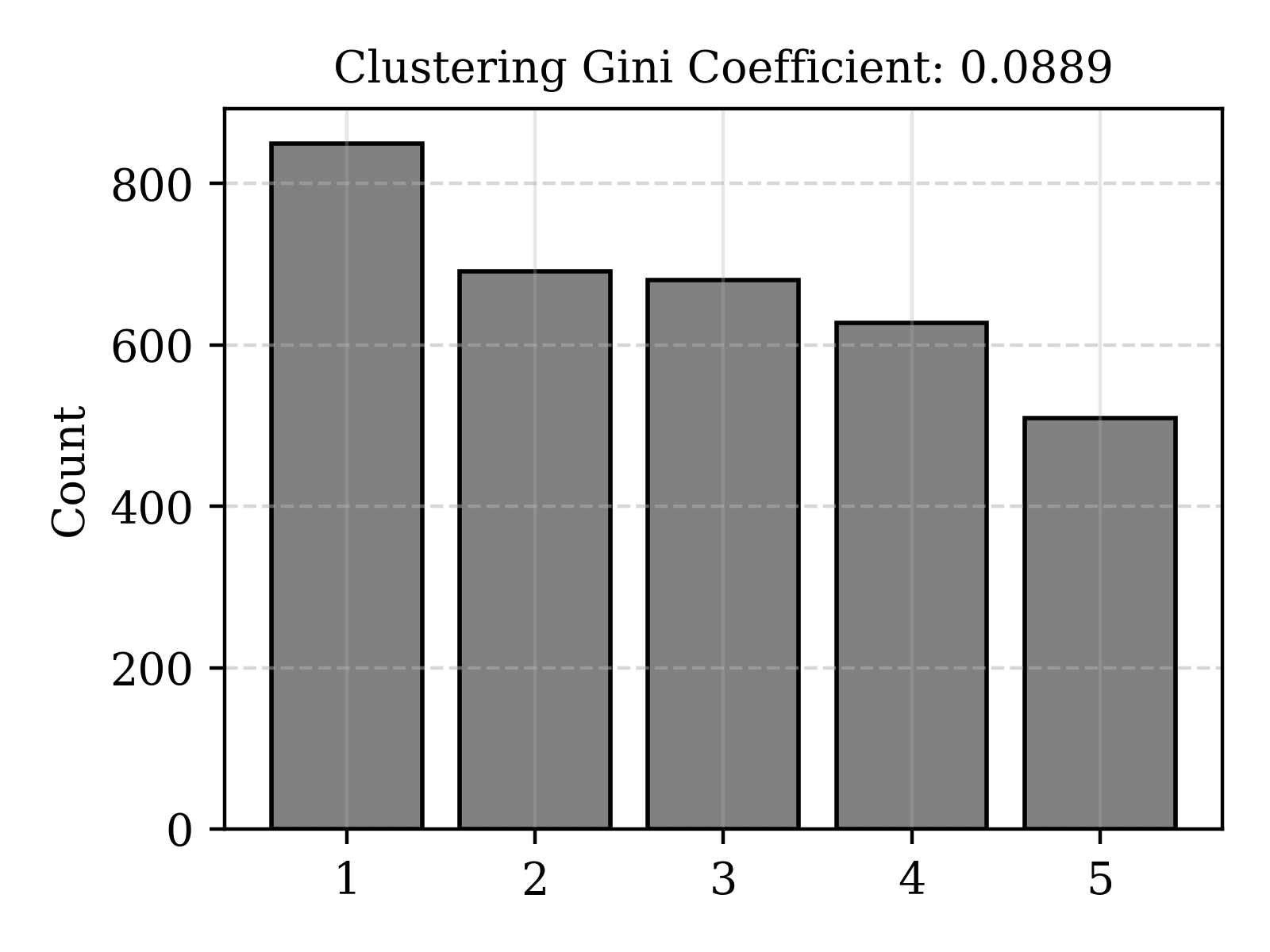}
                    \caption{$k$-Shape clustering ($k=5$).}
                    \label{fig:kshape_cluster_counts}
                \end{subfigure}
                ~
                % Subfigure 2
                \begin{subfigure}[b]{0.45\textwidth}
                    \centering
                    \includegraphics[width=\textwidth, trim={0cm 0cm 0cm 0cm}]{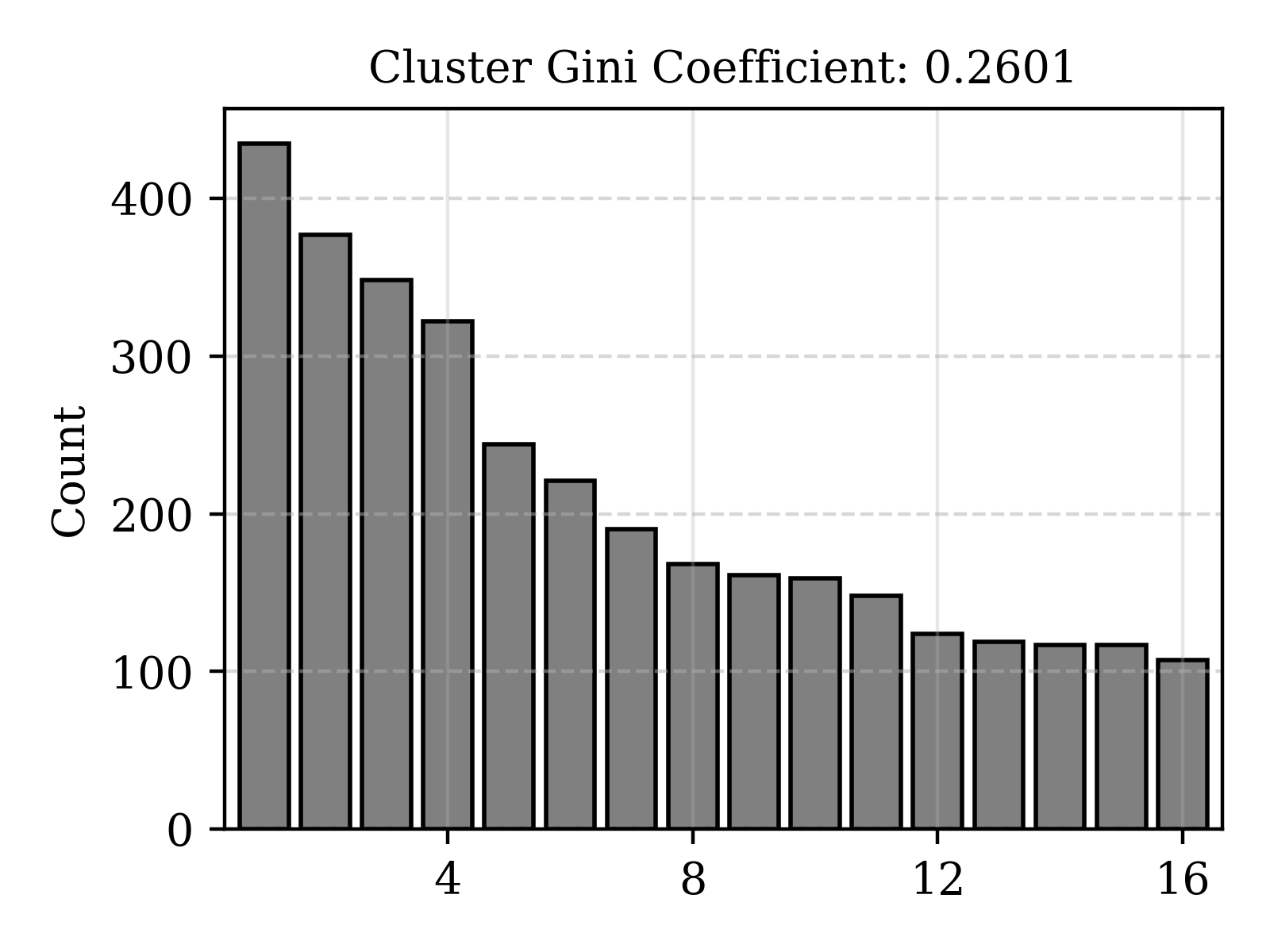}
                    \caption{SOM clustering ($k=16$).}
                    \label{fig:som_cluster_counts}
                \end{subfigure}
                \\
                \vspace{0.5em}
                % Subfigure 3
                \begin{subfigure}[b]{0.45\textwidth}
                    \centering
                    \includegraphics[width=\textwidth, trim={0cm 0cm 0cm 0cm}]{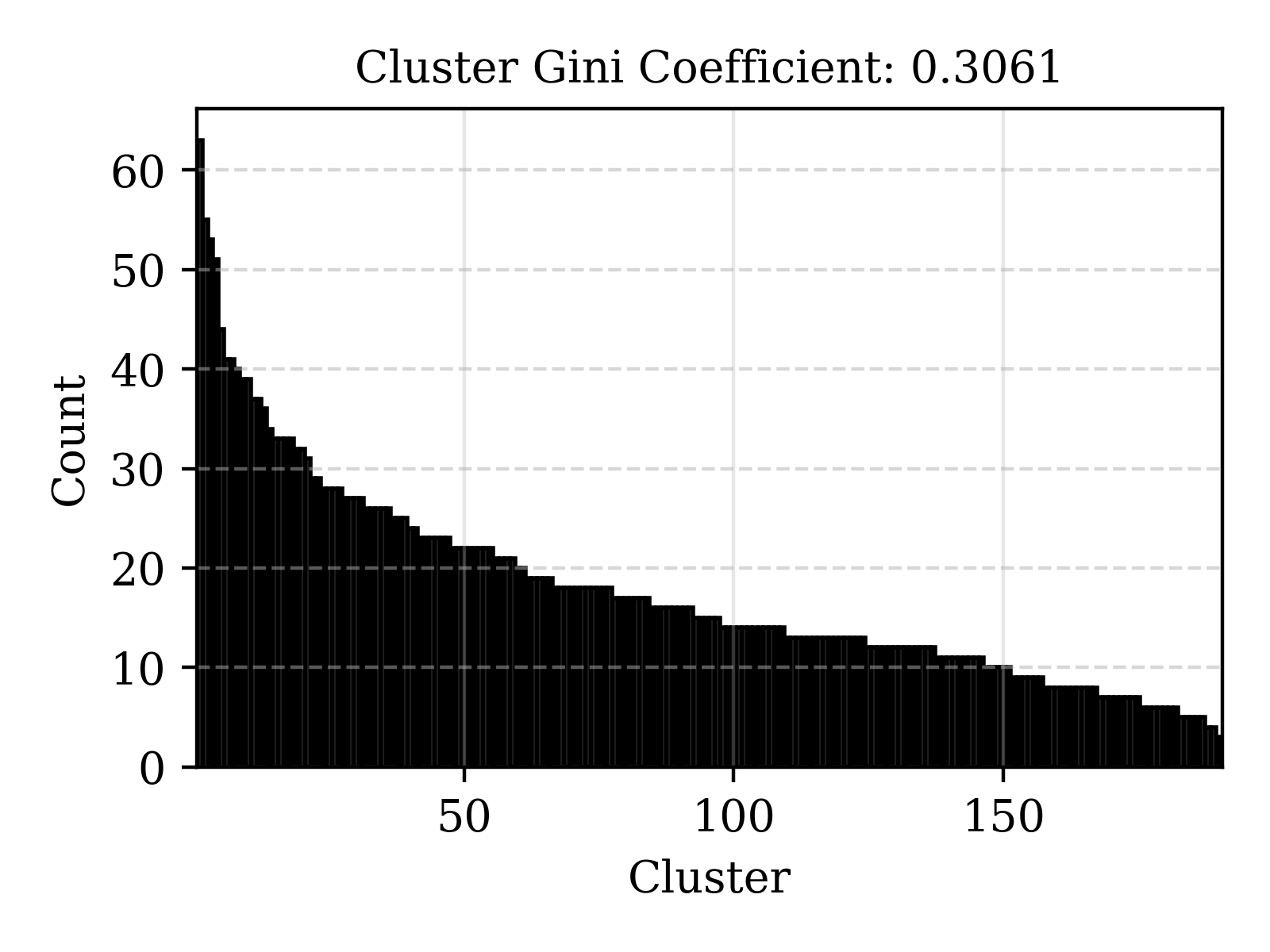}
                    \caption{SBD with AP ($k=190$).}
                    \label{fig:sbd_cluster_counts}
                \end{subfigure}
                ~
                % Subfigure 4
                \begin{subfigure}[b]{0.45\textwidth}
                    \centering
                    \includegraphics[width=\textwidth, trim={0cm 0cm 0cm 0cm}]{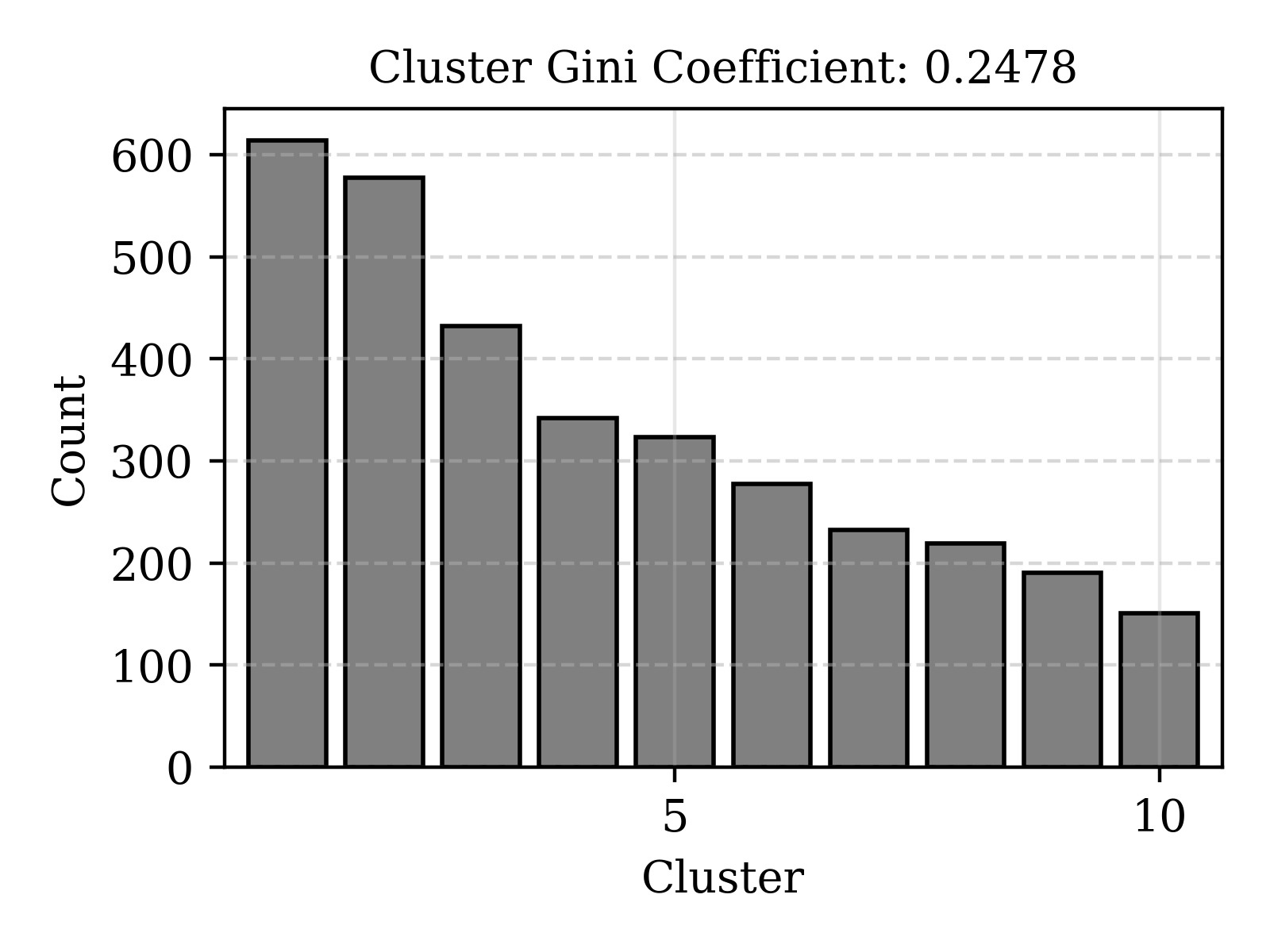}
                    \caption{DBA with GMM ($k=10$).}
                    \label{fig:gmm_cluster_counts}
                \end{subfigure}
                % \vspace{0.5em}
                \caption{Distribution of O\&D pairs across clusters for selected methods at different levels of granularity.}
                \label{fig:cluster_counts}
            \end{figure}

            Figure \ref{fig:cluster_counts} illustrates the distribution of O\&D pairs across clusters for selected methods at various levels of granularity. Methods producing fewer clusters, such as $k$-Shape with $k=5$, tend to exhibit more balanced distributions, suggesting that these broad clusters capture general patterns such as increasing, decreasing, or stable trends across a wide range of O\&D pairs. In contrast, methods like SBD with AP yield a highly uneven distribution of O\&D pairs across clusters, potentially indicating the identification of numerous highly specific patterns that apply to only a small subset of O\&D pairs. Therefore, neither a balanced nor an imbalanced distribution is inherently superior. A balanced distribution may imply that clusters represent overarching marketing influences or airline strategies affecting many O\&D pairs, while a more skewed distribution may indicate rare cases or outliers that warrant further investigation.

        \subsubsection{Clustering Performance Metrics}
        While clustering visualizations provide an intuitive insights, quantitative performance metrics provide additional perspectives on cluster quality. Table \ref{tab:clustering_metrics} presents several metrics that provide information on different aspects of cluster quality such as cohesion, separation, and overall cluster validity.

        \begin{table}[ht!]
            \centering
            \caption{Clustering performance metrics for various methods}
            \label{tab:clustering_metrics}
            \begin{threeparttable}
            \begin{tabular}{p{3cm}lcccccc}
            \toprule
            \textbf{Method\tnote{*}} & \textbf{Clusters} & \textbf{Norm.} & \textbf{Silhouette} & \textbf{D-B Index} & \textbf{Dunn} & \textbf{C-H Index} \\
            \midrule
            \multirow{4}{=}{\raggedright HC with DynTW} 
            & 5 & No & 0.3808 & 1.0381 & 0.0003 & 6698 \\
            & 5 (3)\tnote{\dag} & Yes & -0.0033 & 1.1611 & 0.0145 & 1.599 \\
            & 10 & No & 0.3105 & 1.4915 & 0.0003 & 4637 \\
            & 10 (8)\tnote{\dag} & Yes & -0.1017 & 1.0141 & 0.0145 & 1.278 \\
            \midrule
            \multirow{4}{=}{\raggedright $k$-shape} 
            & 5 (4)\tnote{\dag} & No & 0.2009 & 1.7100 & 0.0002 & 112.5 \\
            & 5 & Yes & 0.0253 & 5.9077 & 0.0146 & 263.7 \\
            & 10 (4)\tnote{\dag} & No & 0.2598 & 5.0568 & 0.0002 & 637.7 \\
            & 10 & Yes & -0.0186 & 5.4534 & 0.0149 & 162.1 \\
            \midrule
            \multirow{4}{=}{\raggedright SOM} 
            & $4 \times 4 = 16$ & No & 0.2408 & 1.4558 & 0.0005 & 4356 \\
            & $4 \times 4 = 16$ & Yes & 0.0594 & 2.3383 & 0.0161 & 238.0 \\
            & $10 \times 10 = 100$ & No & 0.1034 & 1.9382 & 0.0005 & 1098 \\
            & $10 \times 10 = 100$ & Yes & 0.0290 & 2.3517 & 0.0180 & 59.94 \\
            \midrule
            \multirow{4}{=}{\raggedright SBD with AP (+ HC)} 
            & 190 & No & -0.5450 & 14.9283 & 0.0002 & 7.487 \\
            & 190 & Yes & -0.0817 & 3.8756 & 0.0152 & 19.57 \\
            & 190 $\rightarrow$ 10\tnote{\ddag} & No & -0.3184 & 14.4125 & 0.0002 & 20.46 \\
            & 190 $\rightarrow$ 10\tnote{\ddag} & Yes & 0.0314 & 4.2555 & 0.0147 & 128.8 \\
            \midrule
            \multirow{4}{=}{\raggedright DBA with GMM} 
            & 10 & No & -0.0100 & 4.2442 & 0.0002 & 21.36 \\
            & 10 & Yes & -0.0541 & 3.6490 & 0.0148 & 171.1 \\
            & 30 $\rightarrow$ 10\tnote{\ddag} & No & -0.2693 & 6.7018 & 0.0002 & 82.25 \\
            & 30 $\rightarrow$ 10\tnote{\ddag} & Yes & -0.0400 & 3.4532 & 0.0148 & 185.9 \\
            \bottomrule
            \end{tabular}
            \begin{tablenotes}
            \footnotesize
            \item[\dag] Specified number of clusters with the actual number in parentheses due to clustering limitations.
            \item[\ddag] Two-step clustering process, initial clustering followed by further clustering.
            \item[*] \textbf{HC:} Hierarchical Clustering; \textbf{DynTW:} Dynamic Time Warping; \textbf{SOM:} Self-Organizing Maps; \textbf{SBD:} Shape-Based Distance; \textbf{AP:} Affinity Propagation; \textbf{DBA:} DynTW Barycenter Averaging; \textbf{GMM:} Gaussian Mixture Models; %\textbf{Abbreviations:}  Matrix Profile; \textbf{Abbreviations:}  Motif Discovery; \textbf{NMF:} Non-negative Matrix Factorization;
            \textbf{Norm.:} Normalized; \textbf{D-B Index:} Davies-Bouldin Index; \textbf{C-H Index:} Calinski-Harabasz Index.
            \end{tablenotes}
            \end{threeparttable}
        \end{table}

        As mentioned in Section \ref{sec:lr_model}, the Silhouette Coefficient measures how similar an object is to its own cluster compared to other clusters. HC with DynTW without normalization achieved the highest Silhouette score of 0.3808 with 5 clusters, indicating that the clusters were well-separated and compact. In contrast, the Silhouette score decreased significantly when the data were standardized, indicating that normalization affected cluster cohesion and separation by only focusing on the trends rather than magnitude. Additionally, DBA with GMM showed negative Silhouette scores, suggesting that certain O\&D time series patterns may not conform well to Gaussian distributions. The D-B index, which assesses average similarity between each cluster and its nearest neighbor cluster, varied across methods, though, HC with DynTW also had the lowest values, i.e., most distinct clusters. The C-H index, which measures the ratio of the between-cluster dispersion to the within-cluster dispersion, also reinforced these observations. HC with DynTW (without normalization) achieved the highest value of 6698.78, pointing to strong between-cluster dispersion. Furthermore, two-step methods, such as SBD with AP followed by one more step HC, produced higher C-H scores compared to single-step alone, possibly due to initial coarse segmentation followed by refined aggregation.

        On the other hand, the normalized data generally yielded a higher Dunn Index across methods, such as the SOM with 16 neurons achieving a Dunn Index of 0.0161 with normalized data and 0.0005 without normalization. The Dunn Index measures the compactness and separation of clusters by comparing the smallest distance among clusters to the largest intra-cluster distance. Therefore, the results suggested that the normalization tended to have dense and well-separated boundaries of clusters when focusing on the shape of O\&D pairs' local share time series. Overall, each metric offered unique insights. Combining these quantitative evaluations along with visual diagnostics and domain-specific interpretations became essential for a more balanced and comprehensive evaluation.
            
        \subsubsection{Discussion and Challenges}
        Several challenges and limitations were encountered during the clustering analysis. A primary challenge was specifying the number of clusters in advance, as the choice of $k$ was mostly guided by exploratory analysis and domain knowledge. However, the data may not always naturally partition into the specified number of clusters, highlighting the inherent complexity and diversity of O\&D pairs' temporal characteristics. Patterns may exhibit local similarities but diverge across specific temporal periods. Therefore, the introduction of additional feature variables can enhance the comparative analysis of clustering. Moreover, clustering methods such as AP generated a large number of micro-clusters, necessitating the need for subsequent consolidation to enable meaningful interpretation. Despite these challenges, parallel processing and visualization techniques effectively supported the analysis, illustrating that computational intensity and cluster interpretability can be managed through careful method selection and parameter tuning steps. Furthermore, normalization emerged as a beneficial strategy despite the inherent 0-to-1 range of local share data. By prioritizing normalized trends over absolute magnitudes, analysis could detect synchronized behaviors across diverse O\&D pairs. This underscores the potential of normalization as a powerful preprocessing step for reconciling disparate scales within local share data.

\section{Conclusion}\label{sec:Conclusion}

    \subsection{Conclusions from Analysis}
    Local share has emerged as a critical metric for examining passenger composition, offering deep insights into the evolving dynamics of market growth, competition, and airline strategies. System-wide expansion in the airline industry, driven by economic growth, increased passenger demand, and the proliferation of LCCs and ULCCs, has generally elevated local share, reflecting the shifts in market structure and consumer preferences.
    
    The complexity intensifies as market competition intensity, airline mergers, hubbing and de-habbing practice, market entries and exits, and evolving carrier strategies collectively influence local share trends. The resulting fluctuations reveal how market-specific factors, ranging from macroeconomic conditions to localized operational decisions, are reflected in passenger composition patterns. Studying the dynamics in these patterns provides valuable insights into industry trends, competitive landscapes, and strategic behaviors, facilitating a more nuanced understanding of how local share relates to the evolving nature of the airline industry.
    
    \subsection{Key Findings from Modeling}
    The clustering-based modeling provided a multi-faceted understanding of local share dynamics by examining both the magnitude and shape of local share time series patterns across O\&D pairs. Visualization of the clustering outcome offered valuable qualitative insights. The resulting clusters using the original magnitude data tended to reflect distinct levels of local share. Airports serving as major hubs consistently exhibited lower local share values, while large metropolitan coastal airports with strong local demand tended to consistently exhibit higher local share values. Increasing the number of clusters under this setting provided finer hierarchical distinctions, but primarily along the magnitude dimension. In contrast, analyses based on standardized local share patterns placed greater emphasis on the shape and temporal progression of trends rather than the absolute magnitude. Such processing identified pattern commonalities within O\&D pairs, such as sustained growth, steady declines, or plateau phases, even if they differed sustainably in their baseline local share values.

    From a system-level perspective, methods like SOM with standardization effectively captured shape-based patterns across the network, while SBD with AP using up to 190 clusters offered highly granular detail. Hierarchical clustering (HC) also proved valuable in revealing large-scale trend stratification. Furthermore, difficulties in interpreting the clustering results were recurring themes: whereas magnitude-focused methods produced easily understood strata of local share levels, shape-focused processing required careful contextual analysis to explain why certain O\&D pairs followed similar trajectories. Integrating insights from both perspectives, supported by performance metrics and domain knowledge, can lead to a more comprehensive understanding of the factors driving changes in local share within the airline industry.

    \subsection{Potential Implementation and Next Steps}
    Future research directions and practical implementations can build upon the insights gained from the cases analyzed and the clustering results. Improving the interpretability of the clustering results is essential. Incorporating additional attributes, such as detailed airline operational data, route-level pricing information, and airport-specific geographic or socioeconomic factors, could enhance cluster explainability and provide a richer context for the identified patterns. Domain-specific knowledge and expert input, including infrastructure developments, or exogenous shocks (e.g., pandemics or regulatory changes), can further offer deeper explanatory power. Additionally, exploring more advanced time series clustering methods, such as Toeplitz Inverse Covariance-Based Clustering~\cite{hallac2017toeplitz,zhang2021development}, may be beneficial for handling newly introduced attributes and high-dimensional time series data. Integrating well-established clustering methods like AdaBoost~\cite{zhang2023improved} could also leverage the strengths of ensemble learning to enhance clustering performance. More broadly, adopting dynamic and adaptive clustering frameworks may capture evolving market conditions over time, helping to guide strategic decision-making for airlines and policymakers. Further, develop O\&D local share forecasting models to provide insights for long-term planning.    

\bibliographystyle{elsarticle-num} 
\bibliography{LocalShareSubmitV1}




% %% For numbered reference style

%   \textit{\LaTeX: a document preparation system},
%   Addison Wesley, Massachusetts,

% \end{thebibliography}

% \bibliography{sample}

\end{document}